\documentclass[aps,superscriptaddress,onecolumn,a4paper,11pt,reprint]{revtex4}
\usepackage[colorlinks=true, pdfstartview=FitV, linkcolor=blue, citecolor=red, urlcolor=black]{hyperref}
\usepackage{graphicx}
\usepackage[all]{xy}
\usepackage{amsmath}
\usepackage{amssymb}
\usepackage{tensor}
\usepackage{orcidlink}
\newcommand{\be}{\begin{equation}}
\newcommand{\ee}{\end{equation}}
\newcommand{\ben}{\begin{eqnarray}}
\newcommand{\een}{\end{eqnarray}}
\newcommand{\bes}{\begin{subequations}}
\newcommand{\ees}{\end{subequations}}
\def\bal#1\eal{\begin{align}#1\end{align}}

\newcommand{\bfi}{\begin{figure}}
\newcommand{\efi}{\end{figure}}
\newcommand{\bc}{\begin{center}}
\newcommand{\ec}{\end{center}}

\newcommand{\sech}{{\rm sech}}
\newcommand{\p}{{\partial}}

\newcommand{\Nc}{{\cal N}}
\newcommand{\Uc}{{\cal U}}
\newcommand{\Ec}{{\cal E}}
\newcommand{\Ic}{{\cal I}}

\begin{document}

\title{Kink solutions in nonlocal scalar field theory models}
\author{I. Andrade\,\orcidlink{0000-0002-9790-684X}}
        \email[]{andradesigor0@gmail.com}\affiliation{Departamento de F\'\i sica, Universidade Federal da Para\'\i ba, 58051-970 Jo\~ao Pessoa, PB, Brazil}
\author{R. Menezes\,\orcidlink{0000-0002-9586-4308}}
     \email[]{rmenezes@dcx.ufpb.br}\affiliation{Departamento de Ci\^encias Exatas, Universidade Federal
da Para\'{\i}ba, 58297-000 Rio Tinto, PB, Brazil}
\affiliation{Departamento de F\'{\i}sica, Universidade Federal de Campina Grande,  58109-970 Campina Grande, PB, Brazil}
\author{A.Yu. Petrov\,\orcidlink{0000-0003-4516-655X}}
        \email[]{petrov@fisica.ufpb.br}\affiliation{Departamento de F\'\i sica, Universidade Federal da Para\'\i ba, 58051-970 Jo\~ao Pessoa, PB, Brazil}
\author{P.J. Porf\'\i rio\,\orcidlink{0000-0002-4713-3332}}\email[]{pporfirio@fisica.ufpb.br}\affiliation{Departamento de F\'\i sica, Universidade Federal da Para\'\i ba, 58051-970 Jo\~ao Pessoa, PB, Brazil}

\begin{abstract}
In this paper, we study in detail various solutions, especially kink ones, in different nonlocal scalar field theories, whose kinetic term is described by an arbitrary non-polynomial analytic function of the d'Alembertian operator, and the potential is chosen either to be quadratic or to allow for the kink-like solution. Using the perturbative method, we find corrections of first and second orders in the nonlocality parameter around local solutions for several form factors and generate analytic expressions for the energy density up to the first order in this parameter. Additionally, we also address an inverse problem, that is, we reconstruct the potential corresponding to the given solution obtaining restrictions for the form factor.

\end{abstract} 

\maketitle

\section{Introduction}
Nonlocal field theories have been originally motivated by considering finite-size objects within the elementary particles context \cite{efimov}. Further, the concept of nonlocal dynamics has been introduced within other field theory models, especially, gravity, where namely nonlocal extensions are treated as one of the most sound candidates to solve the notorious problem of development of consistent gravity theory. Indeed, as it is well known, the usual Einstein-Hilbert gravity is non-renormalizable, but its higher-derivative extensions which involve higher-curvature geometrical invariants, like $R^2$, $R_{\mu\nu}R^{\mu\nu}$ (the invariant $R_{\mu\nu\alpha\beta}R^{\mu\nu\alpha\beta}$ can be disregarded in four dimensions by using the Gauss-Bonnet topological invariant, ${\cal G}=R^2-4R_{\mu\nu}R^{\mu\nu}+R_{\mu\nu\alpha\beta}R^{\mu\nu\alpha\beta}$), or, more generically, $f(R, R_{\mu\nu}R^{\mu\nu}, R_{\mu\nu\alpha\beta}R^{\mu\nu\alpha\beta})$
, display, in general, the arising of ghosts \cite{Stelle1,Stelle2,buchbinder,frpq} which makes the perturbative treating of these theories highly problematic  (we note that the agravity-like theories \cite{Salvio, Lehum}, whose classical action does not include the usual Einstein-Hilbert term \cite{rodape}, are classically free of this problem, but they display another unpleasant effect, that is, the absence of the consistent Einstein-Hilbert limit). We note that the presence of ghosts is typical for usual, Lorentz invariant higher-derivative field theory models. At the same time, using nonlocal field theories allows us to avoid this problem (see f.e. \cite{Tomboulis};  we note that the nonlocal theories have been argued to be ghost-free also in \cite{Buoninfante00,Barvinsky,Barvinsky2}; the discussion of unitarity in nonlocal theories with use of the Kallen-Lehmann framework was performed in \cite{Nardelli}).  It is worth mentioning that the first attempts of introducing the idea of nonlocality, as far as we know, in field theory, were made over the 1950s, see f.e. \cite{Pais, Pauli, Hay}. The concept of nonlocality also appears in the context of low-energy effective string theory, where infinite-order derivative operators emerge from the perturbative corrections in the $\alpha^{\prime}$  parameter characterizing the inverse of the string tension \cite{Kaku, Green, Witten} (some issues related to string motivations for nonlocality are also presented in \cite{Calcagni:2008nm,Calcagni:2009tx}).

The nonlocality is incorporated in the
quadratic Lagrangian of a real scalar field through ${\cal L}=\frac{1}{2}\phi F(\Box) \phi$, with $F(\Box)$ is a so-called entire function, i.e., that one which does not admit expansion in primitive multipliers like $\prod\limits_{i=1}^n(\Box-M^2_i)^{m_i}$, where all $m_i$ are the integer numbers with $m_i\geq 1$, such a theory is clearly ghost-free. The simplest examples of entire functions are, evidently, trigonometric and exponential ones. Here it must be emphasized that within our paper we restrict the impact of nonlocality by the presence of $F(\Box)$ form factors and do not consider other nonlocal frameworks, such as, for example, Moyal product-based space-time noncommutativity, see f.e. \cite{SW}, and delta function-based nonlocality \cite{Chaichian}). An appropriate choice of $F(\Box)$ allows for the improvement of the renormalizability of a corresponding theory. In the context of gauge theories, the connection between local commutative gauge theories with nonlocal non-commutative ones has been investigated in \cite{Heredia}. Nevertheless, the most prominent application of the nonlocal framework is within gravity since, as it is well known, the Einstein gravity is non-renormalizable, and its usual higher-derivative extensions like $R^2$ gravity involve ghosts. So, first examples of possible nonlocal ghost-free gravity models have been presented in \cite{Biswas,Modesto1,Mazumdar1}, and in \cite{Modesto1}, a first discussion of renormalization properties of such theories has been carried out. Further, the studies of UV behavior of nonlocal gravity have been performed in \cite{Modesto:2012ys,Modesto2,Modesto3} and many other papers. Besides this, nonlocal gravity models are also studied from the classical viewpoint, and some exact solutions for these theories are obtained, especially, within the cosmological context (see f.e. \cite{Calcagni:2009dg,Briscese:2012ys,Biswas,Biswas2}). Apart from cosmology, consistency of various other metrics (f.e. (anti) de Sitter space-time, black holes, G\"{o}del-type solutions) has been checked within nonlocal gravity models \cite{Modesto3a,ourNL,Zhao:2023tox}, Some issues related to degrees of freedom in nonlocal gravity were discussed in \cite{Calcagni:2018gke,Calcagni:2018lyd}. An excellent review on nonlocal gravity is presented in \cite{Modesto4} (more discussions on different aspects of nonlocal gravity also can be found in \cite{nonlocal2022}).

However, studies of non-gravitational field theories within the nonlocal framework are also important. First, it is worth mentioning again the seminal paper \cite{efimov},  where such theories have been applied within the elementary particles context. Second, some perturbative studies of nonlocal field theories have been carried out, see f.e. \cite{Briscese,our2}, and their importance consists not only in the fact that they can allow for phenomenological applications, and, for certain form factors, display all-loop finiteness, but also in their use as a prototype for perturbative calculations in nonlocal gravity. Therefore, the natural question consists in the possibility of obtaining, in analogy with gravity, exact solutions of nonlocal and nonlinear equations of motion, with a perspective of studies of possible topological solutions, thus generalizing results of various studies of kinks and domain walls (see f.e. \cite{vilenkin,manton,vachaspati,trilogia1,Ghoshal:2022mnj,kink1,kink2}), and continuing the research line emphasized in \cite{nonlocaldefects}.  

Within this paper, we obtain some solutions with nontrivial topological characteristics. Particularly, our attention is focused on the study of domain walls, that is, two-dimensional objects that depend on only one spatial coordinate \cite{vachaspati}. Domain walls are the simplest type of localized structures, connecting two adjacent, disconnected minima. As stated in \cite{nonlocaldefects}, topological defects can arise in theories beyond the Standard Model when symmetry is spontaneously broken at some typical high-energy scale. In that work,  for the first time topological defects for nonlocal field theory were investigated, and various interesting results for the domain wall situation were obtained. The corrections to the local solution up to the first order in $1/\Lambda^2$ (high energy scale parameter)  were found with the use of perturbative techniques. It was argued, on the base on analytic calculations,  that, for a specific form factor, this method is only valid near the center of the solution ($x\approx 0$) and far from it ($x\to \infty$). There is an intermediate region where the first-order contribution fails to apply. Additionally, it is observed that the domain wall becomes thicker and has lower energy compared to the local case. Here, we extend the results of \cite{nonlocaldefects}, considering the analysis up to the second order in $1/\Lambda^2$, with a general form factor.

The structure of the paper looks as follows. In section 2, we describe the nonlocal scalar model, derive corresponding equations of motion, and describe the procedure for their solution. In section 3, we obtain some examples of solutions. 
Finally, in section 4 we present our conclusions. Some details of calculations are given in Appendix A.

\section{Model}
Let us start by considering the action of the nonlocal real scalar field in a four-dimensional Minkowski spacetime
\be\label{model}
S = \int d^4x\,\left(\frac12\phi\Box f(\Box_\Lambda)\phi -V(\phi)\right).
\ee
The function $f(\Box_\Lambda)$ is known as the form factor, which depends on the operator $\Box_\Lambda = \Box/\Lambda^2$, with $\Box=\eta^{\mu\nu}\p_\mu\p_\nu$ being the d'Alembertian operator and $\Lambda$ represents a typical high energy scale \cite{Tomboulis}. The potential $V(\phi)$ is a function of the scalar field $\phi$. As usual \cite{Biswas}, the form factor is  assumed to be an entire function,  which can be expanded in power series of $\Box_\Lambda$:
\be\label{fexp}
f(\Box_\Lambda) = \sum_{j=0}^{\infty}c_j\left(\frac{\Box}{\Lambda^2}\right)^j,
\ee
and we fix $c_{0}=1$ to reproduce the correct infrared behavior.

Varying the action \eqref{model} with respect to $\phi$, we obtain the equation of motion for the scalar field
\be\label{feom}
\Box f(\Box_\Lambda)\phi = V_\phi,
\ee
where $V_\phi=dV/d\phi$. The energy-momentum tensor $T_{\mu\nu}$ has been determined in \cite{highorder},
\be\label{Tmunu}
\begin{aligned}
T_{\mu\nu} &= \eta_{\mu\nu}\left(\frac12\phi\Box f(\Box_\Lambda)\phi -V(\phi)\right) +f(\Box_\Lambda)\p_\mu\phi\p_\nu\phi -f(\Box_\Lambda)\phi\p_\mu\p_\nu\phi\\
    &+\frac12\sum_{j=1}^{\infty}\frac{c_j}{\Lambda^{2j}}\sum_{l=1}^{j}\left(\Box^{j-l}\p^\alpha\phi\Box^{l-1}\p_\alpha\p_\mu\p_\nu\phi -\Box^{j-l}\phi\Box^{l}\p_\mu\p_\nu\phi\right),
\end{aligned}
\ee
which is conserved, i.e. $\p_\mu T^{\mu\nu}=0$ as expected. To illustrate, we have for the first terms
\be
\begin{aligned}
T_{\mu\nu} &= \eta_{\mu\nu}\left(\frac12\phi\Box f(\Box_\Lambda)\phi -V(\phi)\right) +\frac{c_0}{2}\left(\p_\mu\phi\p_\nu\phi -\phi\p_\mu\p_\nu\phi\right)\\
    &+\frac{c_1}{2\Lambda^2}\big(\p_\mu\Box\phi\p_\nu\phi -\Box\phi\p_\mu\p_\nu\phi +\p^\alpha\phi\p_\alpha\p_\mu\p_\nu\phi -\phi\p_\mu\p_\nu\Box\phi\big)\\
    &+\frac{c_2}{2\Lambda^4}\big(\p_\mu\Box^2\phi\p_\nu\phi -\Box^2\phi\p_\mu\p_\nu\phi +\p^\alpha\Box\phi\p_\alpha\p_\mu\p_\nu\phi\\
    &-\Box\phi\p_\mu\p_\nu\Box\phi +\p^\alpha\phi\p_\alpha\p_\mu\p_\nu\Box\phi -\phi\p_\mu\p_\nu\Box^2\phi\big) +\mathcal{O}\left(\frac{1}{\Lambda^6}\right).
\end{aligned}
\ee

We now consider solutions that depend only on one spatial dimension, i.e., $\phi=\phi(x)$. In this case, by using Eq. \eqref{fexp}, the equation of motion \eqref{feom} becomes
\be\label{xeom}
\sum_{j=0}^\infty\frac{c_j}{\Lambda^{2j}}\frac{d^{2(j+1)}}{dx^{2(j+1)}}\phi = V_\phi.
\ee
The energy density is given by the component $T_{00}$ of Eq.~\eqref{Tmunu} equal to 
\be\label{rho}
\rho(x) = -\frac12\phi\Box f(\Box_\Lambda)\phi +V(\phi) = -\frac12\phi V_\phi +V(\phi),
\ee
where we have used the equation of motion \eqref{feom}. The energy per unit area can be obtained by integration
\be
E = \int_{-\infty}^{+\infty}dx\,\rho(x).
\ee

For $f=1$, we recover the canonical local model which does not depend on the scale $\Lambda$, as expected. In this situation, Eq.~\eqref{feom} reduces to $\Box\phi = V_\phi$. As a first example, we shall take $V(\phi)=(m^2/2)\phi^2$ which leads to the well-known Klein-Gordon equation, $(\Box-m^2)\phi=0$. For a static one-dimensional field, we have the solution
\be\label{KGsol}
\phi(x) = A\exp(mx) +B\exp(-mx),
\ee
where $A$ and $B$ are real constants. This solution is asymptotically divergent on at least one side, depending on the choice of constants $A$ and $B$. In particular, by the definition of energy density in Eq.~\eqref{rho}, we have $\rho=0$.

We now consider a localized solution connecting two adjacent minima of the potential. These solutions, known as kinks, typically emerge in $Z_2$-symmetric models, see f.e. \cite{Vacha}. As an example, we choose the $\phi^4$ potential whose explicit form looks like
\be\label{phi4}
V(\phi) = \frac{\lambda}{2}\left(v^2 -\phi^2\right)^2,
\ee
where $\lambda$ and $v$ are real positive numbers. This potential presents one local maximum at $\phi=0$ and two minima localized at $\phi=\pm v$. The equation of motion \eqref{xeom} reads
\be\label{eomphi4}
\frac{d^2\phi}{dx^2} = -2\lambda\phi(v^2-\phi^2),
\ee
that supports two non-trivial solutions. The first describes a monotonically increasing function (kink) and the other one describes a monotonically decreasing function (anti-kink). We shall focus on the kink solution:
\be\label{kink}
\phi(x) = v\tanh\big(\sqrt{\lambda}vx\big),
\ee
that connects the $\phi(x\to-\infty)=-v$ to $\phi(x\to+\infty)=+v$. Furthermore, we have conveniently chosen the integration constant to get the solution centered at $x=0$ leading to $\phi(0)=0$. The anti-kink is obtained by changing $x\to-x$. For this solution, the energy density \eqref{rho} becomes
\be\label{rho0}
\rho(x) = \lambda v^4\tanh\big(\sqrt{\lambda}vx\big)^2\sech\big(\sqrt{\lambda}vx\big)^2 +\frac{\lambda v^4}{2}\sech\big(\sqrt{\lambda}vx\big)^4,
\ee
with energy per area $E=4\sqrt{\lambda}v^3/3$.

In general, it is convenient to perform an asymptotic analysis, that is, to consider the case where $x$ is positive and far from the origin. We define $\phi(x)=v-\phi_a(x)$, and substituting it into Eq.~\eqref{eomphi4} considering only linear contributions in $\phi_a$, we have
\be\label{eqasymp}
\frac{d^2\phi_a}{dx^2} = 4\lambda v^2\phi_a.
\ee
The solution of the above equation is the same as Eq.~\eqref{KGsol} with $m^2=4\lambda v^2$. We take $A=0$ since we are interested in solutions that vanish as $x\to\infty$. The constant $B$ is determined by a straightforward comparison with the exact solution \eqref{kink}, then
\be\label{kinkasympt}
\phi_a(x) \approx 2v\exp\big(\!-2\sqrt{\lambda}vx\big).
\ee
We can also analyze the solution around the origin ($x\approx 0$), to obtain
\be\label{kinkori}
\phi(x\approx 0) = \sqrt{\lambda}v^2x -\frac13\sqrt{\lambda^3}v^4x^3 +\frac{2}{15}\sqrt{\lambda^5}v^6x^5 +\mathcal{O}\big(x^7\big).
\ee
Up to now, we treated only local cases.
However, note that the results we obtained are crucial for guiding subsequent findings, and furthermore, they should be recoverable in the limit $\Lambda\to\infty$.

\section{Nonlocal solutions}

In this section, we shall obtain nonlocal solutions for different kinds of form factors, namely:
\bes\label{f}
\bal
\label{f1}f_{I}(\Box_\Lambda) &=\exp\big(\Box/\Lambda^2\big),\\
\label{f2}f_{II}(\Box_\Lambda) &=\exp\big(-\Box/\Lambda^2\big),\\
\label{f3}f_{III}(\Box_\Lambda) &= \cosh\big(\sqrt{\Box/\Lambda^2}\big),\\
\label{f4}f_{IV}(\Box_\Lambda) &= \cos\big(\sqrt{\Box/\Lambda^2}\big).
\eal
\ees
We note that these form factors certainly do not exhaust all possibilities. A more generic discussion of nonlocal form factors can be found in \cite{nonlocal2022}.

\subsection{Nonlocal Klein-Gordon equation}
We start by considering the potential $V(\phi)=(m^2/2)\phi^2$, which leads us to a nonlocal generalization of the Klein-Gordon equation, $\Box f(\Box_\Lambda)\phi = m^2\phi$. As we are only interested in static one-dimensional solutions, we must solve Eq.~\eqref{xeom}. In this case, we write
\be\label{xeomKG}
\sum_{j=0}^\infty\frac{c_j}{\Lambda^{2j}}\frac{d^{2(j+1)}}{dx^{2(j+1)}}\phi = m^2\phi.
\ee
Although the equation involves infinite-order derivatives, it is linear in $\phi$. Inspired by Eq.~\eqref{KGsol}, we suggest  an ansatz for a solution of the following form
\be\label{solM}
\phi(x) = A\exp(Mx) +B\exp(-Mx).
\ee
 This is a natural ansatz for a solution of a linear differential equation with constant coefficients, with $M$ are roots of the corresponding characteristic equation we write below. In general, these roots are complex and hence do not display a desired asymptotics (monotonous decay as $x\to\pm\infty$), so, for the example of a desired behavior, we restrict ourselves to a case of one real positive root $M$.
The $M$ parameter, depending on $\Lambda$, controls the growth/decay of the static field. Substituting the above solution into Eq.~\eqref{xeomKG}, we obtain the characteristic equation
\be\label{eqM}
M^2\sum_{j=0}^\infty\frac{c_jM^{2j}}{\Lambda^{2j}} = m^2.
\ee
From this algebraic equation, we determine $M$. We can do it either by analytical or by numerical methods.  We require the $M$ to recover the known behavior $M=m$ at $\Lambda\to\infty$. To do so, it is necessary to provide the explicit form of $f(\Box_\Lambda)$. Since the nonlocality scale is required to be large, $\Lambda^2/m^2\gg 1$, it is possible to expand Eq.~\eqref{solM} in power series
\be
\begin{aligned}
\phi(x) &= A\exp(mx) +B\exp(-mx) -\frac{c_1mx}{2\Lambda^2}\left(A\exp(mx) -B\exp(-mx)\right)\\
    &+\frac{m^5x}{8\Lambda^4}\left(4c_2 -c_1^2\left(7-mx\right)\right)\left(A\exp(mx) +B\exp(-mx)\right) +\mathcal{O}\left(\frac{1}{\Lambda^6}\right).
\end{aligned}
\ee
As we want to evaluate a solution that vanishes as $x\to\infty$, we set $A=0$ and rewrite the above equation considering a higher order in the expansion:
\be\label{Pasy}
\phi(x) = B\exp(-mx)\left(1 +\frac{c_1m^3x}{2\Lambda^2} +\frac{m^5x}{8\Lambda^4}\left(4c_2 -c_1^2\left(7-mx\right)\right)\right) +\mathcal{O}\left(\frac{1}{\Lambda^6}\right).
\ee
In principle, the constant $B$ is arbitrary and may depend on $\Lambda$, since the limit $\Lambda \to \infty$ is respected.

The importance of studying the nonlocal Klein-Gordon equation \eqref{xeomKG} is that it allows us to understand the asymptotic behavior of $\phi_+(x)=v-\phi_a(x)$ for the non-linear situation given by \eqref{xeom} since this potential can be expanded near the minimum, $\langle\phi\rangle=v$, as $V(\phi)\approx\mu\phi^2+\mathcal{O}(\phi^3)$, with $2\mu=V_{\phi\phi}|_{\phi=v}$. We continue our discussion by considering examples of $f(\Box_\Lambda)$.

\begin{figure}[htb!]
    \centering
    \includegraphics[width=5.4cm]{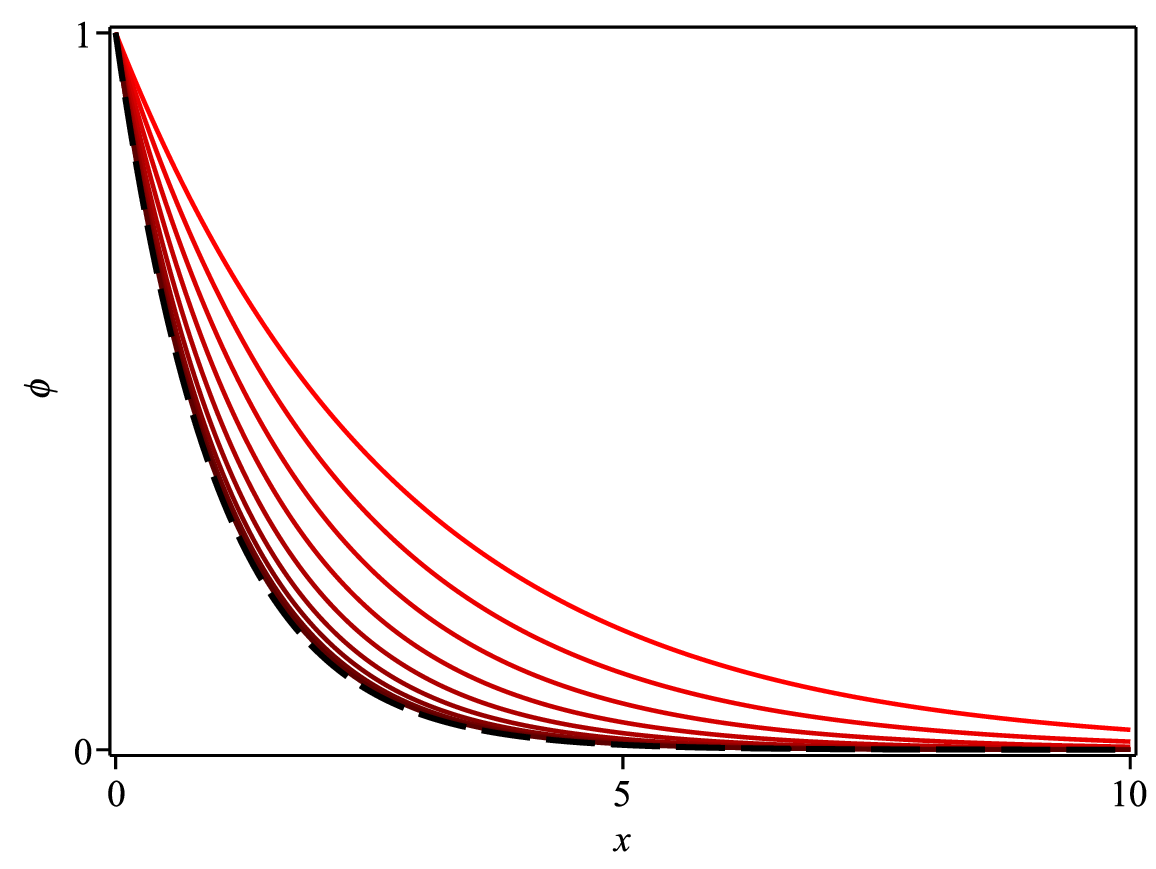}
    \includegraphics[width=5.4cm]{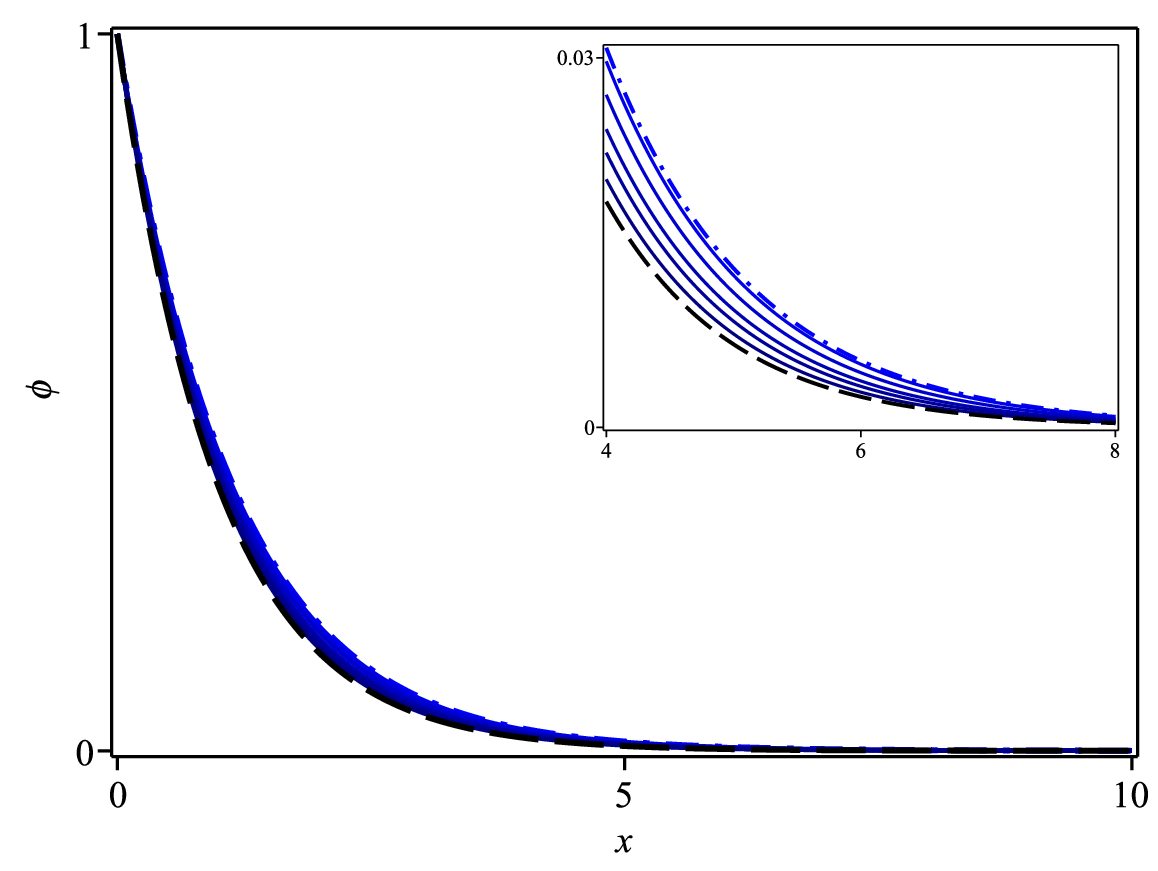}
    \includegraphics[width=5.4cm]{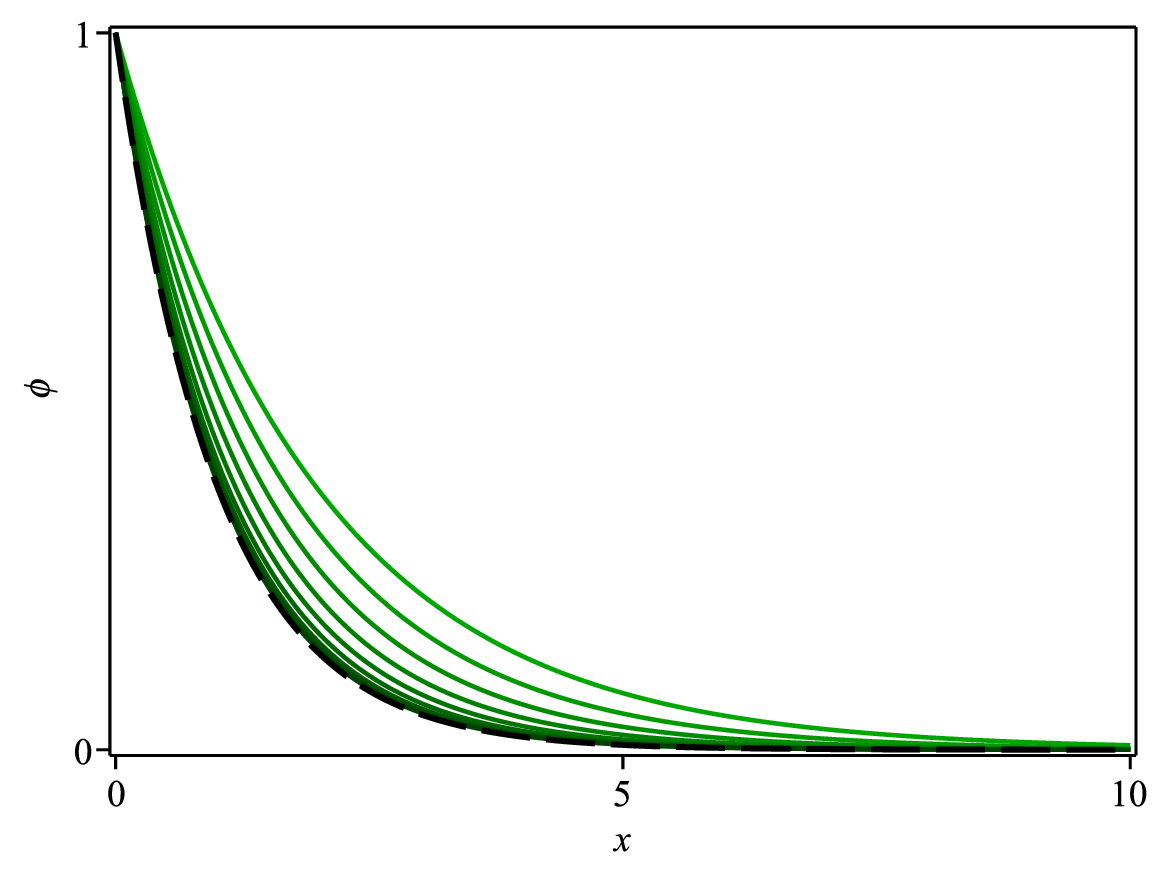}
    \caption{The solutions $\phi(x)$ associated to Eqs.~\eqref{KGsol+} (left) and \eqref{KGsol-} (middle), and for $\phi(x)=\exp(-Mx)$ with $M$ numerically obtained from Eq.~\eqref{Mcosh} (right). We set $A=0$ and $m=B=1$. On the left and right panels, the values of $\Lambda^2$ are $\Lambda^2=2^i$ with $i=-4..4$, varying from red/green to black as $\Lambda^2$ increases. For the middle sketch, the values are as follows: $\Lambda^2=\exp(1)$ for the dash-dotted line, and for the solid lines, $\Lambda^2=3,4,6,9,20$, with colors transitioning from blue to black as $\Lambda^2$ increases. The dashed line represents $\Lambda^2\to\infty$ in all plots and is shown in black.}
    \label{fig1}
\end{figure}

As a first example, we consider the form factor given by Eq.~\eqref{f1}, where the coefficients are $c_j=\frac{1}{j!}$. The Eq.~\eqref{eqM} becomes
\be
M^2\exp\big(M^2/\Lambda^2\big) =  m^2,
\ee
whose solution is $M=\sqrt{W_0(m^2/\Lambda^2)}\Lambda$, where $W_0(y)$ is the principal branch of Lambert $W$ function with argument $y$, with $y\geq -\exp(-1)$. The solution in Eq.~\eqref{solM} reads
\be\label{KGsol+}
\phi(x) = A\exp\big(\sqrt{W_0(m^2/\Lambda^2)}\Lambda x\big) +B\exp\big(-\sqrt{W_0(m^2/\Lambda^2)}\Lambda x\big).
\ee
Taking $\Lambda\to\infty$, we recover the solution in Eq.~\eqref{KGsol}, as expected. The decay/growth parameter of this solution depends on $\Lambda$. To ensure that the solution vanishes as $x$ approaches infinity, we take $A=0$. In this case, we can use the expansion \eqref{Pasy} to write
\be\label{asymptfpos}
\phi(x) = B\exp(-mx)\left(1 +\frac{m^3x}{2\Lambda^2} -\frac{m^5x}{8\Lambda^4}\left(5-mx\right)\right) +\mathcal{O}\left(\frac{1}{\Lambda^6}\right).
\ee
The behavior of the solution \eqref{KGsol+} is shown in the left panel of Fig.~\ref{fig1}, for $A=0$, $B=m=1$ and several values of $\Lambda$, ranging from the limit as $\Lambda\to\infty$ to some non-zero values.

The second example is similar to the previous one, with the form factor given by Eq.~\eqref{f2}. Here, we have $c_j=(-1)^j/j!$ and the Eq.~\eqref{eqM} turns
\be
M^2\exp\big(-M^2/\Lambda^2\big) =  m^2.
\ee
Thus, the solution $\phi(x)$ is expressed as
\be\label{KGsol-}
\phi(x) = A\exp\big(\sqrt{-W_0(-m^2/\Lambda^2)}\Lambda x\big) +B\exp\big(-\sqrt{-W_0(-m^2/\Lambda^2)}\Lambda x\big).
\ee
Unlike the previous case, for this situation, the value of $\Lambda$ is bounded from below, with $\Lambda\geq\Tilde{\Lambda}$, where $\Tilde{\Lambda}=m\exp(1/2)$. The Eq.~\eqref{Pasy} becomes
\be\label{asymptfneg}
\phi(x) = B\exp(-mx)\left(1 -\frac{m^3x}{2\Lambda^2} -\frac{m^5x}{8\Lambda^4}\left(5-mx\right)\right) +\mathcal{O}\left(\frac{1}{\Lambda^6}\right).
\ee
In the middle panel of Fig.~\ref{fig1}, we plot the solution given by Eq.~\eqref{KGsol-} for $A=0$, $B=m=1$ and several values of $\Lambda$ within the previously established range, $\Lambda\in[\Tilde{\Lambda},\infty)$.

The third example of the form factor is given by Eq.~\eqref{f3}, with $c_j=\frac{1}{(2j)!}$. In this case, the Eq.~\eqref{eqM} reduces to
\be\label{Mcosh}
M^2\cosh\big(\sqrt{M^2/\Lambda^2}\big) =  m^2.
\ee
Unfortunately, we are unable to derive an analytical solution for this equation. However, one can use Eq.~\eqref{Pasy} to find an expression for $\phi(x)$ in power series of $1/\Lambda^2$, 
\be
\phi(x) = B\exp(-mx)\left(1 +\frac{m^3x}{4\Lambda^2} -\frac{m^5x}{96\Lambda^4}\left(19-3mx\right)\right) +\mathcal{O}\left(\frac{1}{\Lambda^6}\right).
\ee
Similarly to the previous examples, we still observe decay behavior for the field. We plotted numerically for $A=0$, $m=B=1$ and different values of $\Lambda$, see right panel of Fig.~\ref{fig1}. The form factor in Eq.~\eqref{f4}, whose $c_j=(-1)^j/(2j)!$, can be also evaluated numerically.

\subsection{Nonlocal Klein-Gordon equation with $\phi^{4}$-potential}
Now we consider the $\phi^4$-potential given by Eq.~\eqref{phi4}. Therefore, the equation of motion \eqref{xeom} takes the form
\be\label{xeomphi4}
\sum_{j=0}^\infty\frac{c_j}{\Lambda^{2j}}\frac{d^{2(j+1)}}{dx^{2(j+1)}}\phi = -2\lambda\phi(v^2- \phi^2),
\ee
which is a nonlinear equation with infinite order in derivatives, making it quite challenging to obtain an analytical solution. However, using the results found above we can determine the behavior of the field near the minimum, $\phi\approx v$, when $x\to\infty$. Substituting $\phi_+(x)=v-\phi_a(x)$ in the above equation, we recover the nonlocal solution \eqref{solM} with $m^2=4\lambda v^2$ and $A=0$. As it is known, this solution decreases, vanishing as $x\to\infty$, providing us with the asymptotic profile $\phi_+(x)$.

As a result of the above asymptotic analysis, the constant $B$ cannot be determined as it is factored in the field equation. Nevertheless, we estimate its value from Eq.~\eqref{kinkasympt}, as the approximation is only valid for high values of $\Lambda$, and for the limit $\Lambda\to\infty$, this value must be recovered. Therefore, we set $B=2v$. 
A more rigorous analysis could be conducted for $B(\Lambda)=2v+\mathcal{O}(1/\Lambda^2)$, as performed in this subsection.

Given the difficulty of solving Eq.~\eqref{xeomphi4} analytically, one can employ the perturbative method \cite{nonlocaldefects} in order to carry out this task. We use some results briefly outlined in the appendix \ref{pertub}. This method involves considering perturbations of $1/\Lambda^2$ around the local solution \eqref{kink}. To calculate the first-order correction, we solve the non-homogeneous ODE in Eq.~\eqref{edo1o}. In our case, this equation becomes
\be
\begin{aligned}
&\frac{d^2\xi_1}{dx^2} -2\lambda v^2\left(2 -3\,\sech^2\big(\sqrt{\lambda}vx\big)\right)\xi_1 = 8c_1\lambda^2v^5\left(1 -3\,\sech^2\big(\sqrt{\lambda}vx\big)\right)\tanh\big(\sqrt{\lambda}vx\big)\sech^2\big(\sqrt{\lambda}vx\big),
\end{aligned}
\ee
with a general solution
\be\label{xi1}
\begin{aligned}
\xi_1(x) &= \Nc_1\,\sech^2\big(\sqrt{\lambda}vx\big) +\Nc_2\left[3\sqrt{\lambda}vx\,\sech^2\big(\sqrt{\lambda}vx\big) +\tanh\big(\sqrt{\lambda}vx\big)\left(3 +2\cosh^2\big(\sqrt{\lambda}vx\big)\right)\right]\\
&+2c_1\lambda v^3\sech^2\big(\sqrt{\lambda}vx\big)\left(2\tanh\big(\sqrt{\lambda}vx\big) -\sqrt{\lambda}vx\right),
\end{aligned}
\ee
where $\Nc_1$ and $\Nc_2$ are integration constants. The first-order perturbation solution is $\phi_1(x) = \phi_0(x) + (1/\Lambda^2)\xi_1(x)$. To ensure that this function resembles the profile of Eq.~\eqref{kink}, we impose some conditions on $\xi_1(x)$. Firstly, the solution must only exist within the interval from $-v$ to $v$, with $\phi_1(x\to\pm\infty)=\pm v$. To achieve this, we require $\xi_1(x\to\pm\infty)=0$, so $\Nc_2=0$. Furthermore, we aim to maintain the solution symmetry, i.e., $\phi_1(x)=-\phi_1(-x)$, which is only possible with $\Nc_1=0$. Resulting in
\be\label{phi1}
\phi_1(x) = v\tanh\big(\sqrt{\lambda}vx\big) +\frac{2c_1\lambda v^3}{\Lambda^2}\sech^2\big(\sqrt{\lambda}vx\big)\left(2\tanh\big(\sqrt{\lambda}vx\big) -\sqrt{\lambda}vx\right).
\ee
The energy density for the above solution is $\rho=\rho_0+(1/\Lambda^2)\rho_1$, where $\rho_0$ is the local energy density given by Eq.~\eqref{rho0} and $\rho_1$ is its the first-order correction, which the detailed computation is presented in Appendix \ref{pertub} and expressed as
\be
\rho_1(x) = -4c_1\lambda^2v^6\tanh^3\big(\sqrt{\lambda}vx\big)\left(2\tanh\big(\sqrt{\lambda}vx\big) -\sqrt{\lambda}vx\right).
\ee
The energy per area of the solution \eqref{phi1} is given by
\be
E_1 = \frac{4\sqrt{\lambda}v^3}{3} -\frac{8c_1\sqrt{\lambda^3}v^5}{15\Lambda^2},
\ee
which for the perturbation scheme to make sense, we must have $c_1\ll 5\Lambda^2/(2\lambda v^2)$.

Despite the aforementioned constraints, the analysis is not yet complete. Depending on the parameter values, the monotonicity condition $\phi_1'(x)>0$ may not be valid. This requirement helps us in deriving the necessary restrictions for the parameters. Examining the solution \eqref{phi1} at $x\approx0$, we obtain
\be\label{phi1ori}
\phi_1(x\approx0) \approx \sqrt{\lambda}v^2\left(1+\frac{2c_1\lambda v^2}{\Lambda^2}\right)x -\frac{\sqrt{\lambda^3}v^4}{3}\left(1+\frac{10c_1\lambda v^2}{\Lambda^2}\right)x^3 +\mathcal{O}(x^5).
\ee
From this, we need $c_1\gg -\Lambda^2/(2\lambda v^2)$. Asymptotically, we are left with
\be\label{phi1asympt}
\phi_1(x) \approx v -2v\exp\big(\!-2\sqrt{\lambda}vx\big)\left(1 -\frac{4c_1\lambda v^2}{\Lambda^2}\left(2-\sqrt{\lambda}vx\right)\right) +\mathcal{O}\left(\exp\big(\!-4\sqrt{\lambda}vx\big)\right).
\ee
This behavior is similar to that of Eq.~\eqref{asymptfpos} with $m^2=4\lambda v^2$ and $B=2v(1-8\lambda v^2/\Lambda^2)$. Continuing the analysis for the monotonicity of this behavior for $x$ far from the origin, it is necessary that $c_1\gg-\Lambda^2/\big(4\big(\sqrt{\lambda}v\big)^3x\big)$. Note that for negative $c_1$, this condition is never valid, regardless of how large the value of $\Lambda$ is, as $x\to\infty$. For this reason, we consider only form factors with positive $c_1$. For $c_1>0$, the condition below Eq.~\eqref{phi1ori} is always satisfied. Under these conditions, we ensure the monotonicity $\phi_1(x)$ around the origin and at $x_{+}$. For the form factors \eqref{f1} and \eqref{f3}, the solution monotonically increases, whereas for the form factors \eqref{f2} and \eqref{f4}, the solution oscillates around the vacuum.

\begin{figure}[htb!]
    \centering
    \includegraphics[width=8cm]{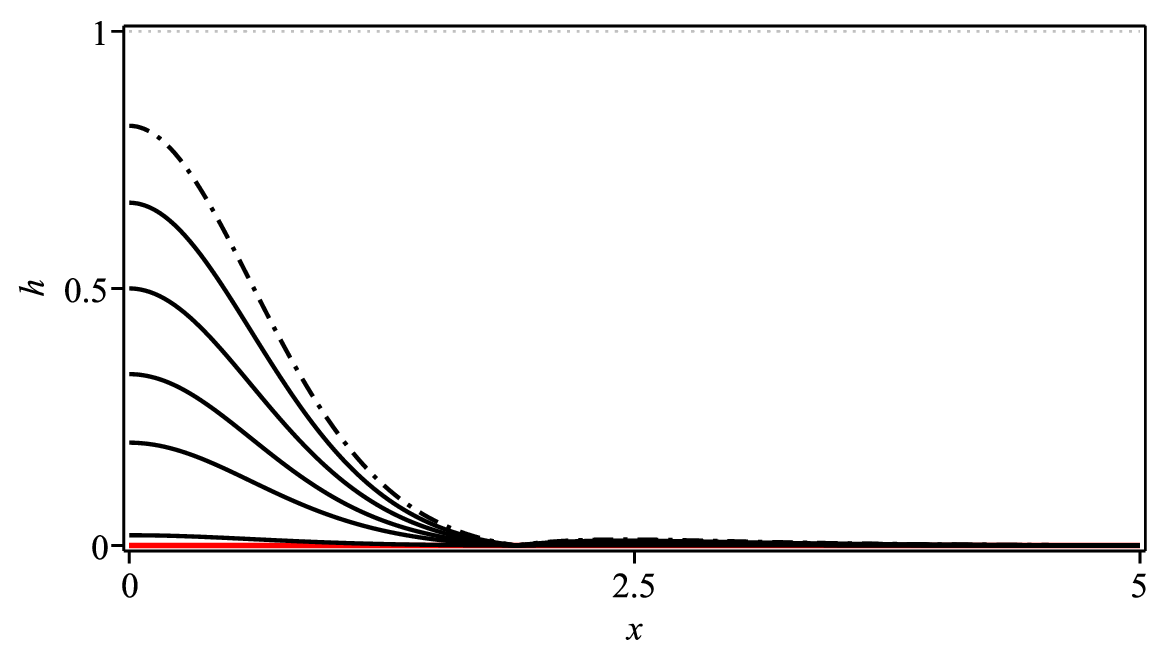}
    \includegraphics[width=8cm]{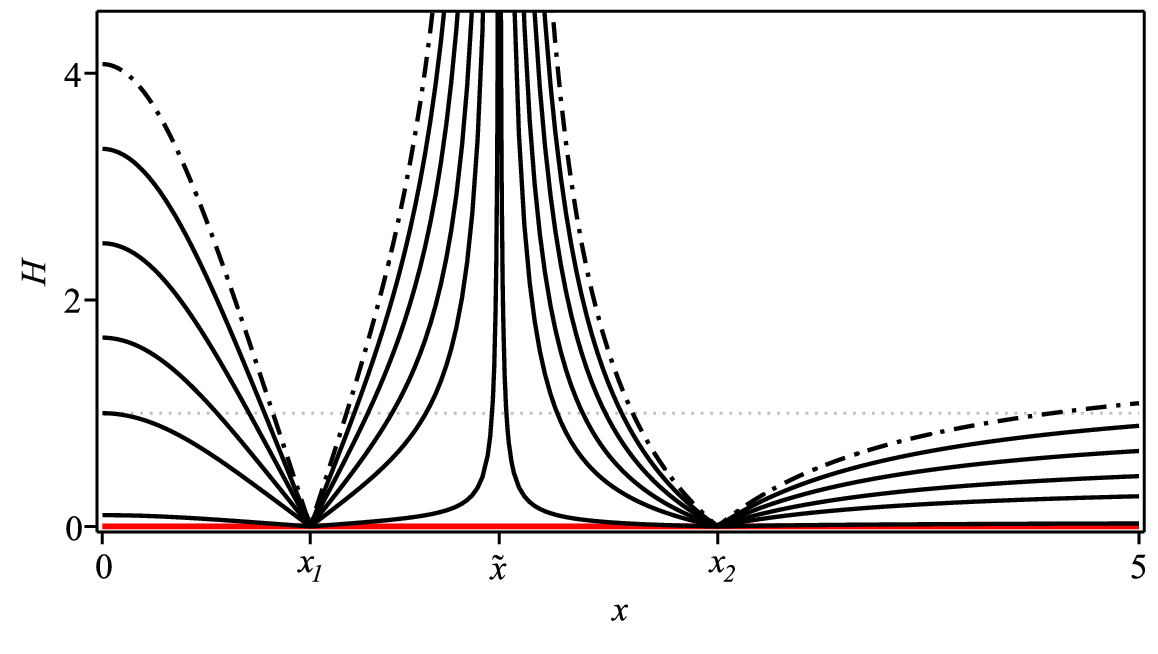}
    \caption{The functions $h(x)$ (left) and $H(x)$ (right) in Eq.~\eqref{hH1} for $\lambda=v=c_1=1$. The dash-dotted line stands for $\Lambda^2=\bar{\Lambda}^2=2.45$, while the solid lines represent $\Lambda^2=3,4,6,10,100$ in black. The red line corresponds to an analysis of the local solution in Eq.~\eqref{kink}, which is given by $h(x)=H(x)=0$ (no perturbation). The values at the tick marks are $x_1=1.002$, $x_2=2.968$, and $\Tilde{x}=1.915$.}
    \label{fig2}
\end{figure}

We analyze in detail only two regions, $x\approx0$ and $x\to\infty$, but there is an intermediate region between them. A more rigorous analysis can be performed by taking the derivative of Eq.~\eqref{phi1} and verifying its positivity, we have
\be
\frac{d\phi_1}{dx} = \sqrt{\lambda}v^2\sech^2\big(\sqrt{\lambda}vx\big)\left(1 +\frac{2c_1\lambda v^2}{\Lambda^2}\left(2\sqrt{\lambda}vx\tanh\big(\sqrt{\lambda}vx\big) +6\,\sech^2\big(\sqrt{\lambda}vx\big) -5\right)\right) > 0.
\ee
Evaluating this condition analytically is intricate, but we can do it numerically. For the form factor \eqref{f1}, with $c_1=\lambda=v=1$, the equation above is fulfilled if $\Lambda^2>\bar{\Lambda
}^2=2.45$. However, this condition does not suffice to guarantee the validity of the perturbation $\xi_1(x)$. We must verify if the requirements in Eq.~\eqref{hH} are satisfied, where
\bes\label{hH1}
\bal
h(x) &= \frac{2v^2\lambda|c_1|}{\Lambda^2}\left|\frac{2\tanh\big(\sqrt{\lambda}vx\big) -\sqrt{\lambda}vx}{\tanh\big(\sqrt{\lambda}vx\big)}\right| \gg1,\\
\label{hH1b}H(x) &= \frac{2v^2\lambda|c_1|}{\Lambda^2}\left|\frac{6\big(1-2\,\sech^2\big(\sqrt{\lambda}vx\big)\big)\tanh\big(\sqrt{\lambda}vx\big) -\big(2-3\,\sech^2\big(\sqrt{\lambda}vx\big)\big)\sqrt{\lambda}vx}{2\tanh\big(\sqrt{\lambda}vx\big) -\sqrt{\lambda}vx}\right| \gg1.
\eal
\ees
The first requirement is satisfied depending on the chosen value of $\Lambda^2$. On the other hand, the second one presents a divergence at $x=\tilde{x}$ which corresponds to take $2\tanh\big(\sqrt{\lambda}v\tilde{x}\big)=\sqrt{\lambda}v\tilde{x}$ in Eq.~\eqref{hH1b}, it is solved (particularly for $\lambda=v=1$, $\tilde{x}=1.915$). This divergence appears regardless of the chosen form factor. In Fig.~\ref{fig2}, we plot the functions $h(x)$ and $H(x)$ for $\lambda=v=1$ and some values of $\Lambda$, with the form factor \eqref{f1}. The perturbation scheme is only valid near and far from the origin. It is observed in Fig.~\ref{fig2} that in the intermediate region between $x=x_1=1.002$ and $x=x_2=2.968$, the perturbative scheme is broken down. Although our results are obtained for specific parameter values, this behavior of the function $H(x)$ is generic. Such a property has already been observed for a particular form factor \cite{nonlocaldefects}.

We analyze the symmetric and localized perturbation \eqref{xi1}, i.e., $\Nc_1=\Nc_2=0$. The requirement $\Nc_2=0$ is a necessary condition to maintain the smallness of the perturbative character of the solution; at the same time, setting $\Nc_1=0$ was a choice made to preserve the symmetry $\phi_1(x) = -\phi_1(-x)$. In this case, we observe that the function $H(x)$ in Eq.~\eqref{hH1b} exhibits two symmetric divergences. For $\Nc_1 \neq 0$, the perturbed solution $\phi_1(x)$ no longer holds the above symmetry, and in any case, it is not possible to avoid divergences by analyzing the function $H(x)$, defined in \eqref{hH}. Keeping $\Nc_1\neq 0$, it is worth mentioning that this analysis shows that the number of divergences depends on the value of $\Nc_1$, occurring at the values of $\tilde{x}$ where $\Nc_1 + 2c_1\lambda v^3\big(2\tanh\big(\sqrt{\lambda}v\tilde{x}\big) - \sqrt{\lambda}v\tilde{x}\big) = 0$. Specifically, for $\lambda = v = c_1 = 1$, the function $H(x)$ has three divergences for $0 < |\Nc_1| < 1.066$, two for $|\Nc_1| = 1.066$, and only one for $|\Nc_1| > 1.066$. These divergences are not symmetric, unlike the case when $\Nc_1 = 0$. Note that, although it is possible to reduce the number of divergences in $H(x)$, they cannot be completely eliminated, and a divergence arises at $x=0$ appears in the function $h(x)$, which does not exist when $\Nc_1$ is null. In this sense, choosing $\Nc_1=0$ is the most convenient option for studying the first-order correction.

We can extend our study and consider the second-order correction $\xi_2(x)$. This contribution can be obtained from Eq.~\eqref{edo2o}, which, for the potential \eqref{phi4}, is written as
\be\label{edo2otanh}
\begin{aligned}
\frac{d^2\xi_2}{dx^2} +c_1\frac{d^4\xi_1}{dx^4} +c_2\frac{d^6\phi_0}{dx^6} = 6\lambda\phi_0\xi_1^2 -2\lambda\big(v^2-3\phi_0^2\big)\xi_2,
\end{aligned}
\ee
where $\phi_0(x)$ is the local solution of Eq.~\eqref{kink} and $\xi_1(x)$ is the first-order correction given by Eq.~\eqref{xi1} with $\Nc_1=\Nc_2=0$. Solving this equation, one finds
\be
\begin{aligned}
\xi_2(x) &= 2\lambda^2v^5\sech^2\big(\sqrt{\lambda}vx\big)\Big[\left((32c_1^2-15c_2)\tanh\big(\sqrt{\lambda}vx\big) -12c_1^2\sqrt{\lambda}vx\right)\sech^2\big(\sqrt{\lambda}vx\big)\\
    &-2\big((12+\lambda v^2x^2)c_1^2 -5c_2\big)\tanh\big(\sqrt{\lambda}vx\big) +(15c_1^2-4c_2)\sqrt{\lambda}vx\Big].
\end{aligned}
\ee
The perturbed solution can be set in the following form: $\phi_2(x)=\phi_1(x)+(1/\Lambda^4)\xi_2(x)$. As we know, the solution with the first correction $\phi_1$ is not valid throughout space; thus, $\phi_2$ is not either. However, it remains consistent near the origin $x \approx 0$ and far from it, $x \to \infty$. Near the origin, we have
\be\label{phi2ori}
\begin{aligned}
\phi_2(x\approx0) &\approx \sqrt{\lambda}v^2\left(1+\frac{2c_1\lambda v^2}{\Lambda^2} +\frac{2\lambda^2 v^4(11c_1^2-9c_2)}{\Lambda^4}\right)x\\
    &-\frac{\sqrt{\lambda^3}v^4}{3}\left(1+\frac{10c_1\lambda v^2}{\Lambda^2} +\frac{2\lambda^2v^4(107c_1^2-77c_2)}{\Lambda^4}\right)x^3 +\mathcal{O}(x^5),
\end{aligned}
\ee
and asymptotically
\be\label{phi2asympt}
\begin{aligned}
\phi_2(x) &\approx v -2v\exp\big(\!-2\sqrt{\lambda}vx\big)\bigg[1 -\frac{4c_1\lambda v^2}{\Lambda^2}\left(2-\sqrt{\lambda}vx\right)\\
    &+\frac{4\lambda^2v^4}{\Lambda^4}\left(24c_1^2-10c_2 -(15c_1^2-4c_2)\sqrt{\lambda}vx +2c_1^2\lambda v^2x^2\right)\bigg] +\mathcal{O}\left(\exp\big(\!-4\sqrt{\lambda}vx\big)\right).
\end{aligned}
\ee
 The monotonicity of the solution in these regions is ensured by a restriction on $c_2$, which can be easily obtained since $\Lambda$ is large. The validity of the perturbation $\xi_2(x)$ can be checked using the results in the appendix \ref{pertub}. For reasonable values of $\Lambda$, the approximation holds in the intervals $|x|<\bar{x}_1=0.671$ and $|x|>\bar{x}_2=5.5$. It is noteworthy that the region where the perturbation scheme breaks down is enlarged by considering the second-order corrections.

\subsection{Nonlocal Klein-Gordon equation with infinite powers of the scalar field}
In the previous subsection, we addressed the challenge of obtaining a solution for $\phi(x)$. Our initial attempt was to explore a potential for which we have an analytical solution in the local case, such as that of Eq. \eqref{phi4}, and seek a solution for the nonlocal case. However, it is feasible to consider an alternative approach. We can then search for a potential whose solution is given by Eq. \eqref{kink}. In this way, we consider the following potential
\be\label{VU}
V(\phi) = \frac{\lambda}{2}\left(v^2 -\phi^2\right)^2\Uc(\phi,\Lambda),
\ee
where $\Uc(\phi,\Lambda)$ is the deformation function. The case $\Uc(\phi,\Lambda)=1$ recovers the original potential \eqref{phi4}.
This function must be such that the potential $V(\phi)$ retains minimal values at $\phi^2=v^2$ and the local maximum at $\phi=0$. In this case, we choose a deformation function that does not alter the signs of its first derivatives ($V_\phi$, $V_{\phi\phi}$, and so on) between the minima.

Substituting Eqs.~\eqref{kink} and \eqref{VU} in Eq.~\eqref{xeom}, we get
\be\label{Vgeral}
V(\phi) = \frac{\lambda}{2}\left(v^2 -\phi^2\right)^2\sum_{j=0}^\infty c_j\left(\frac{4\lambda v^2}{\Lambda^2}\right)^j\left(\sum_{k=0}^{2j}(-1)^{k}\left(\sum_{l=0}^k g_l(\phi)\right)\left(\sum_{m=k}^{2j}h_m(\phi)\right)\right),
\ee
with
\bes
\bal
g_l(\phi) &= \frac{(-1)^l}{(2v)^l}(l+1)!S(k+1,l+1)(v-\phi)^l,\\
h_m(\phi) &= \frac{(-1)^{m-k}}{(2v)^{m-k}}(m-k+1)!S(2j-k+1,m-k+1)(v-\phi)^{m-k},
\eal
\ees
where $S(a,b)$ is a Stirling number of the second kind with parameters $a$ and $b$. 

Expanding Eq.~\eqref{Vgeral} up to order $1/\Lambda^4$, we see that
\be
V(\phi) = \frac{\lambda}{2}\left(v^2 -\phi^2\right)^2\left(1 +\frac{8c_1\lambda}{\Lambda^2}\left(\phi^2 -\frac{v^2}{2}\right) +\frac{180c_2\lambda^2}{\Lambda^4}\left(\phi^4 -\frac{10v^2\phi^2}{9} +\frac{v^4}{5}\right) +\mathcal{O}\left(\frac{1}{\Lambda^6}\right)\right).
\ee
The higher power of this expansion is $\phi^8$. In general, for expansion of the $1/\Lambda^{2n}$, we get $\phi^{4+2n}$. Therefore, this potential represents an infinite series. The expression for the energy density \eqref{rho} is cumbersome, so we omit it here. However, it is possible to calculate the expansion of the energy per area in orders of $1/\Lambda^2$, yielding
\be\label{energyU}
E = \frac{4\sqrt{\lambda}v^3}{3} -\frac{32c_1\sqrt{\lambda^3}v^5}{15\Lambda^2} +\frac{64c_2\sqrt{\lambda^5}v^7}{7\Lambda^4} -\frac{1024c_3\sqrt{\lambda^7}v^9}{15\Lambda^6} +\frac{25600c_4\sqrt{\lambda^9}v^{11}}{33\Lambda^8} +\mathcal{O}\left(\frac{1}{\Lambda^{10}}\right),
\ee
that depends explicitly of the form factor.

The requirements imposed in the deformation function naturally lead to restrictions on choices for the form factor $f(\Box_\Lambda)$. Since we desire to obtain small deviations from the shape of $\phi^{4}$-potential, we require its height of the local maximum not to depart significantly from $V(0)=\lambda v^4/2$. For $\phi=0$, the potential \eqref{Vgeral} is
\be\label{V0}
V(0) = \frac{\lambda v^4}{2} -\frac{2c_1\lambda^2 v^6}{\Lambda^2} +\frac{18c_2\lambda^3 v^8}{\Lambda^4} -\frac{304c_3\lambda^4 v^{10}}{\Lambda^6} +\frac{8608c_4\lambda^5 v^{12}}{\Lambda^8} -\frac{374016c_5\lambda^6 v^{14}}{\Lambda^{10}} +\mathcal{O}\left(\frac{1}{\Lambda^{12}}\right).
\ee
As can be seen, it is necessary to choose coefficients $c_j$ conveniently so that this expansion does not diverge. We have $V(0)=\sum_jK_j/\Lambda^{2j}$, thus $K_j$ must be smaller than a typical high energy scale. For this, we impose $\lim\limits_{j\to\infty}K_j/\Lambda^{2j}=0$. That is not always possible; we perform this analysis for some $c_j$ already used in this work (form factors in Eq.~\eqref{f}). The energy in Eq.~\eqref{energyU} can also be written as $E=\sum_j\Ec_j/\Lambda^{2j}$, and it is possible to perform a similar analysis to ensure that it does not diverge.

For the form factors \eqref{f1} and \eqref{f2}, with corresponding coefficients $c_j=1/j!$ and $c_j=(-1)^j/j!$, we have
\be\label{V0ex12}
V(0) = \frac{\lambda v^4}{2} \mp\frac{2\lambda^2 v^6}{\Lambda^2} +\frac{9\lambda^3 v^8}{\Lambda^4} \mp\frac{152\lambda^4 v^{10}}{3\Lambda^6} +\frac{1076\lambda^5 v^{12}}{3\Lambda^8} \mp\frac{15584\lambda^6 v^{14}}{5\Lambda^{10}} +\mathcal{O}\left(\frac{1}{\Lambda^{12}}\right).
\ee
The convergence condition $\lim\limits_{j\to\infty}K_j/\Lambda^{2j}=0$ is not valid, regardless of how big is the typical high energy scale, $\Lambda$. To check this out, we computed this limit for $\Lambda^2=10,10^3$, and $10^5$ and found that it diverges. We also checked this limit for the diverged energy \eqref{energyU}.

\begin{figure}[htb!]
    \centering
    \hspace{2mm}\includegraphics[width=7.8cm]{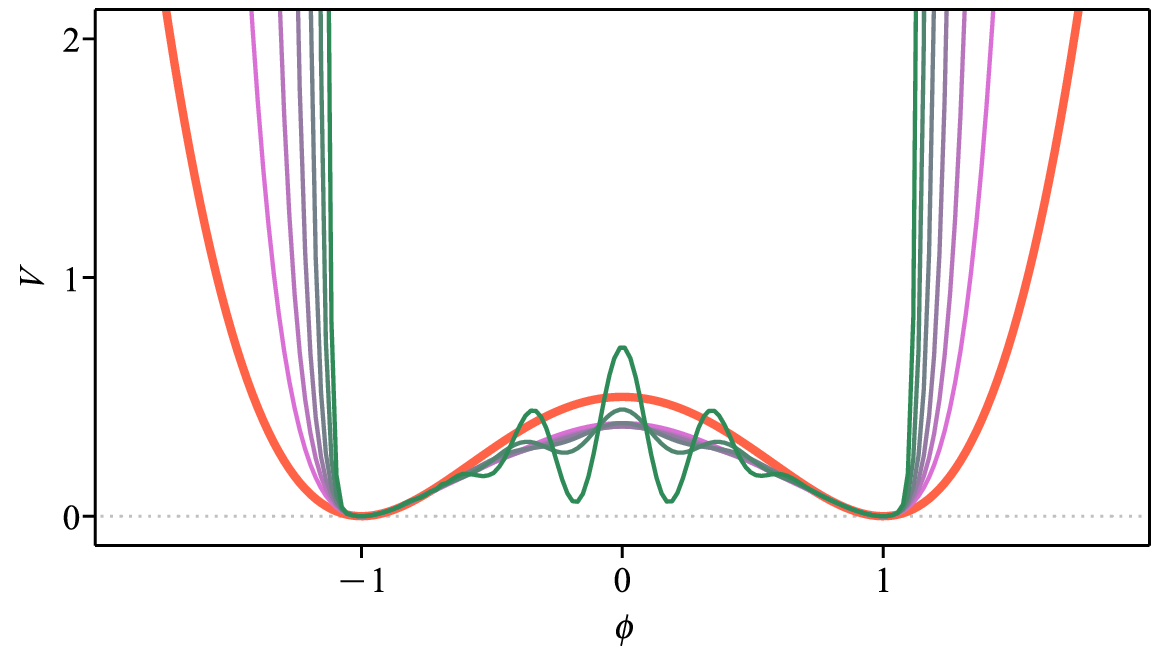}
    \hspace{1mm}\includegraphics[width=7.8cm]{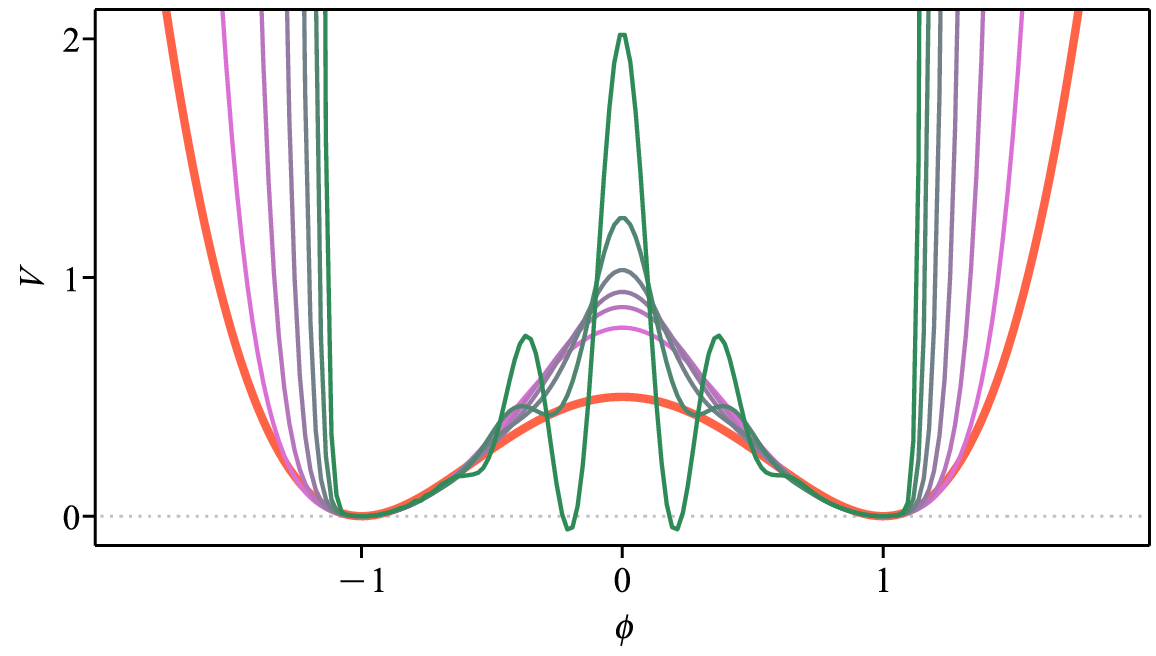}
    \includegraphics[width=8cm]{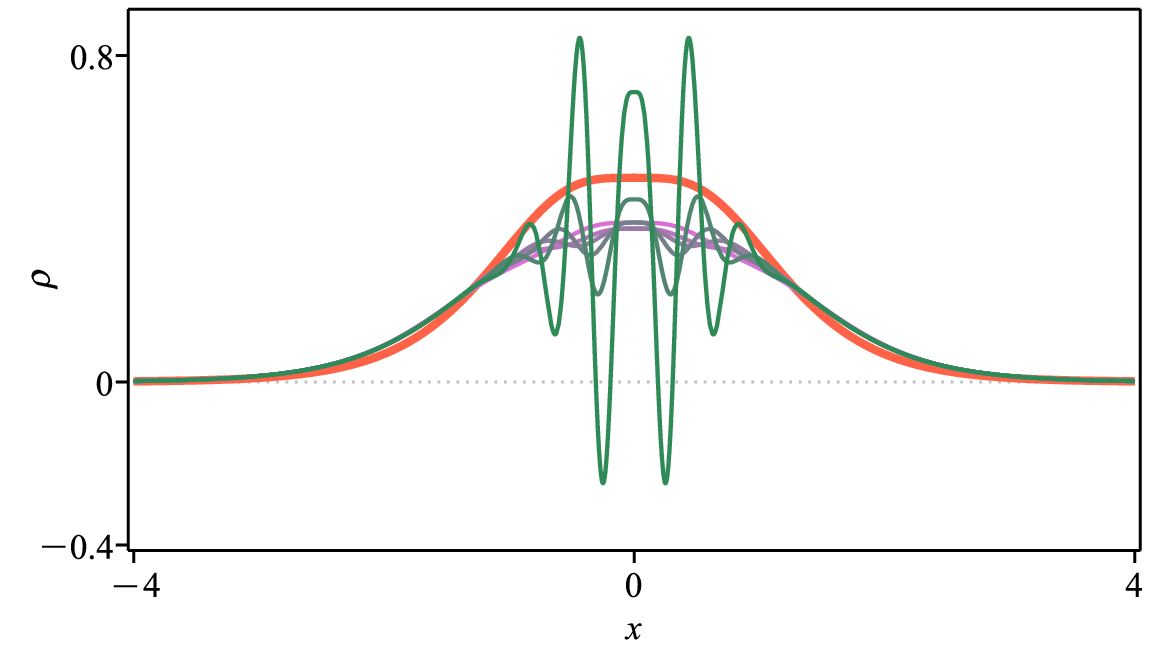}
    \hspace{-1mm}\includegraphics[width=8cm]{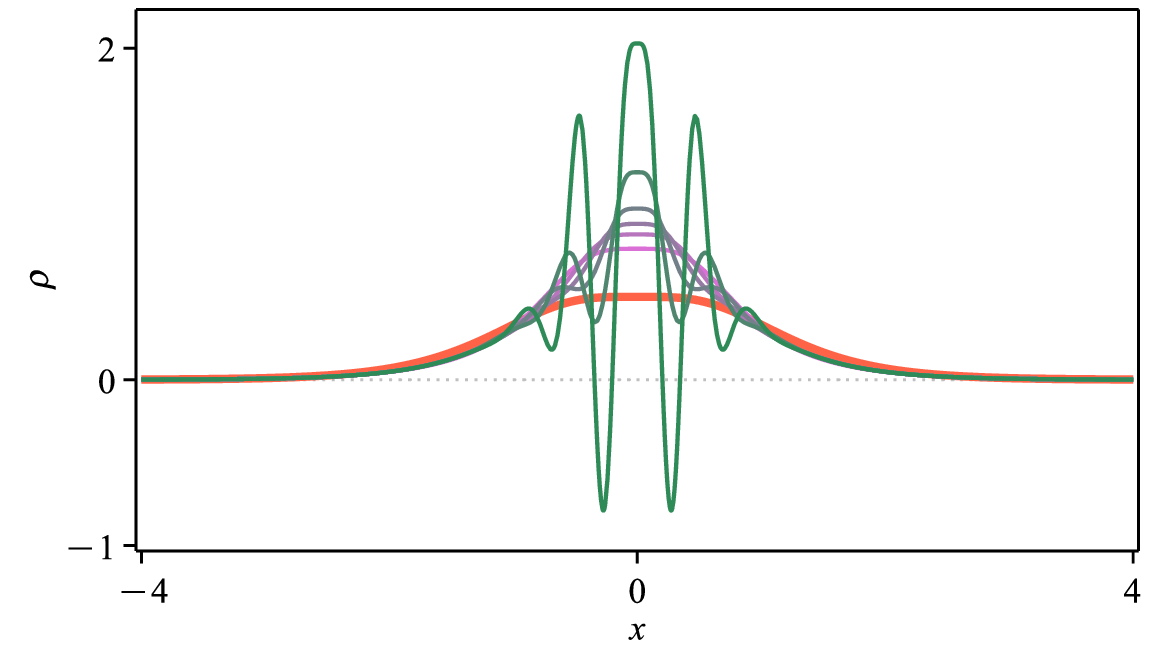}
    \caption{Expansion in powers of $1/\Lambda^2$ of the potential $V(\phi)$ in Eq. \eqref{Vgeral}, the top panels, and energy density for $\rho(x)$ in Eq.~\eqref{rho}, the bottom panels. The left side is for $c_j=1/j!$ and the right side for $c_j=(-1)^j/j!$ for $\lambda=v=1$ and $\Lambda^2=10$. In both, the orange line represents order $0$, recovering \eqref{phi4} and \eqref{rho0}, the other curves represent orders $(1/\Lambda^2)^{2k}$ with $k=1,...,6$ from lilac to green.}
    \label{fig3}
\end{figure}

\begin{figure}[htb!]
    \centering
    \includegraphics[width=8cm]{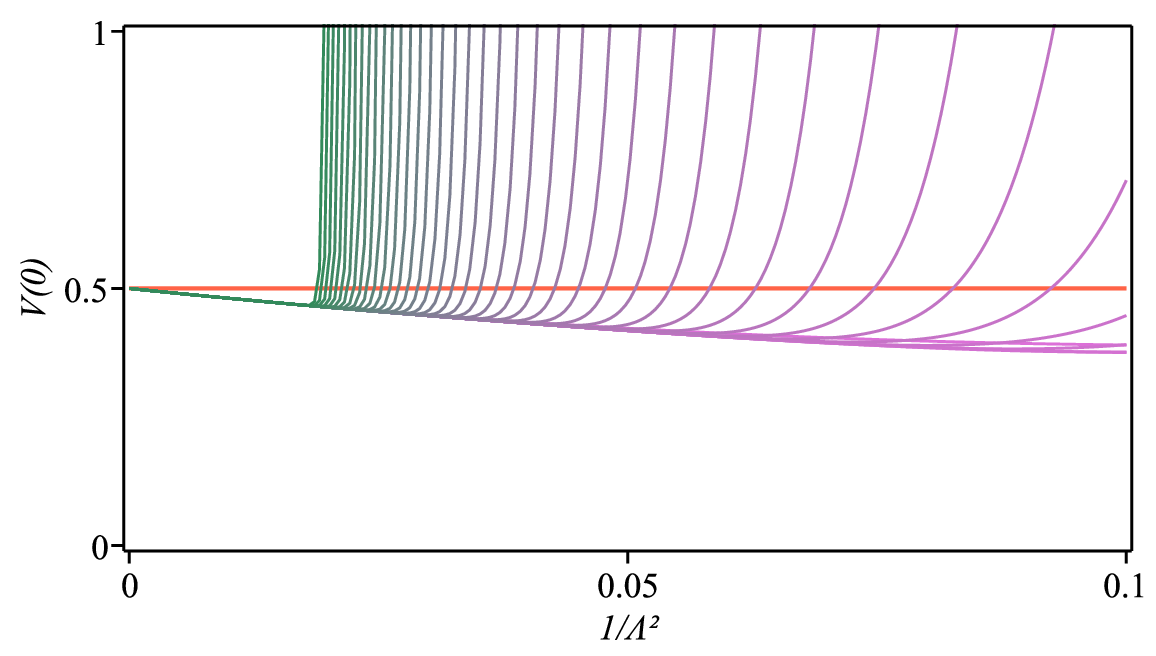}
    \includegraphics[width=8cm]{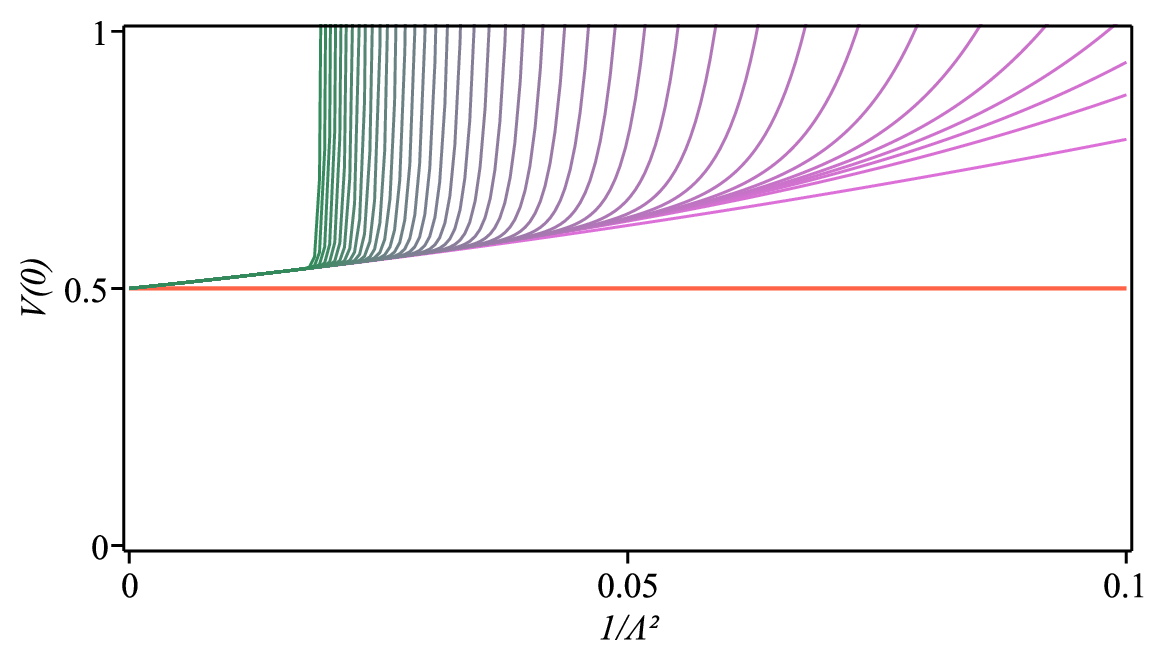}
    \includegraphics[width=8cm]{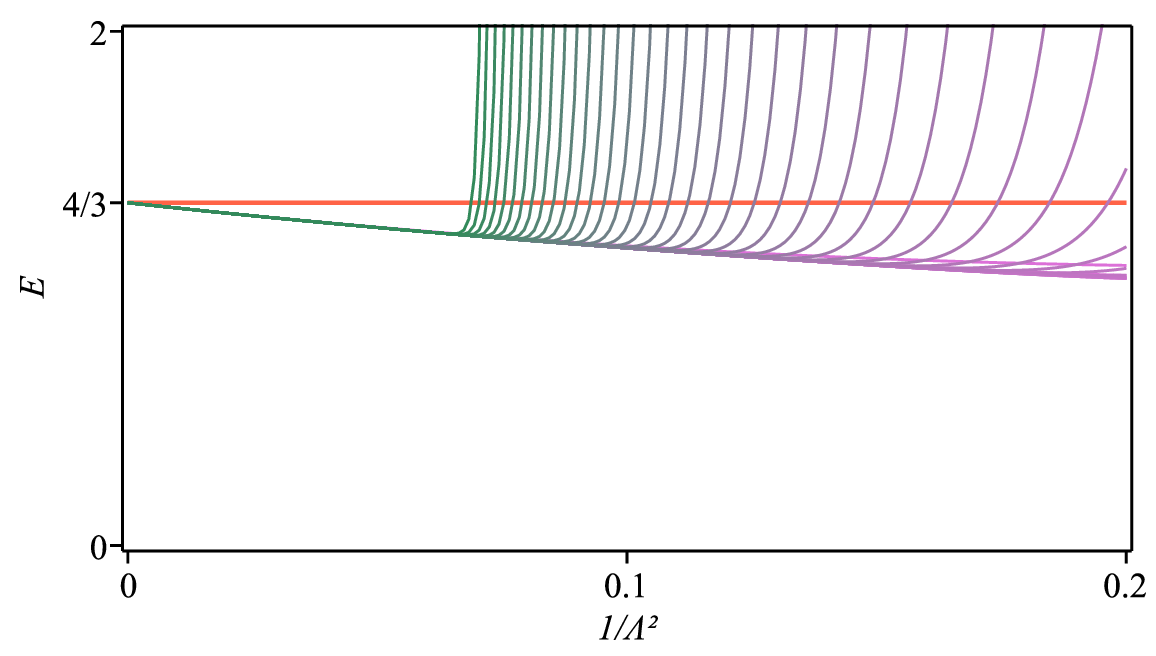}
    \includegraphics[width=8cm]{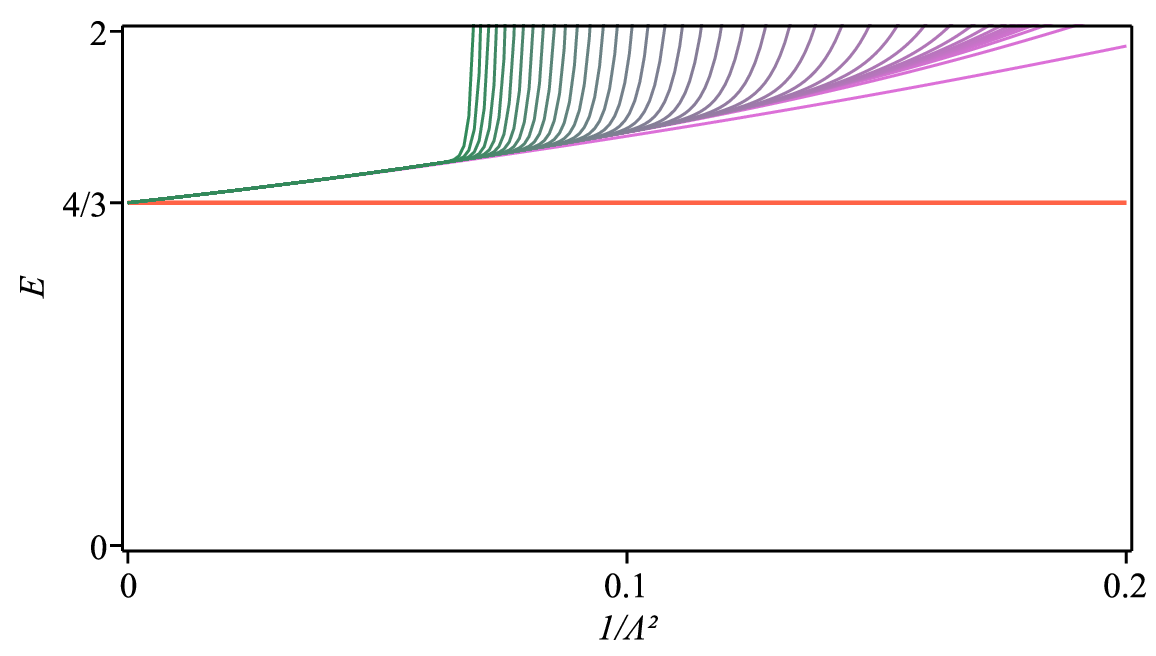}
    \caption{Potential height $V(0)$ (top) obtained from Eq. \eqref{V0ex12} and energy $E$ (bottom) from Eq. \eqref{energyU} in terms of $1/\Lambda^2$. The left panels correspond to the form factor given by \eqref{f1}, while the right panels correspond to \eqref{f2}, both with $\lambda=v=1$. Each line represents orders of $(1/\Lambda^{2})^{2k}$, ranging from 0 to 40. The color of the lines transitions from green to lilac as the order increases from $2$ to $80$ of $1/\Lambda^2$, with the orange line denoting order $0$.}
    \label{fig4}
\end{figure}

In practice, it is not viable to evaluate the sums in the potential \eqref{Vgeral}, therefore, we truncate the expansion at specific orders of $1/\Lambda^2$ to observe the shape of $V(\phi)$ and for the energy density $\rho(x)$. In Fig.~\ref{fig3}, we plot the potential \eqref{Vgeral} and the energy density \eqref{rho} at several orders of $1/\Lambda^2$ for $c_j=1/j!$ and $c_j=(-1)^j/j!$, with $\lambda=v=1$ and $\Lambda^2=10$. Notice that in the initial terms of the expansion, the potential $V(\phi)$ has the desired form. However, as higher orders are considered, its derivatives exhibit multiple sign changes within $-1<\phi<1$, and the maximum at $\phi=0$ undergoes significant variations in its value, as displayed in the top panels of the Fig.~\ref{fig3}. The same situation occurs for the energy density, as shown in the bottom panels of the Fig.~\ref{fig3}. This behavior persists regardless of the typical high energy scale, $\Lambda$. However, it only begins to emerge at higher orders of $1/\Lambda^2$. The profile of the potential height $V(0)$ and the energy $E$, concerning $1/\Lambda^2$, can be observed in Fig.~\ref{fig4}. We display its values for various orders of $1/\Lambda^2$, and it can be noticed that even if $\Lambda$ is large, there is divergence in both at high orders.

\begin{figure}[htb!]
    \centering
    \includegraphics[width=8cm]{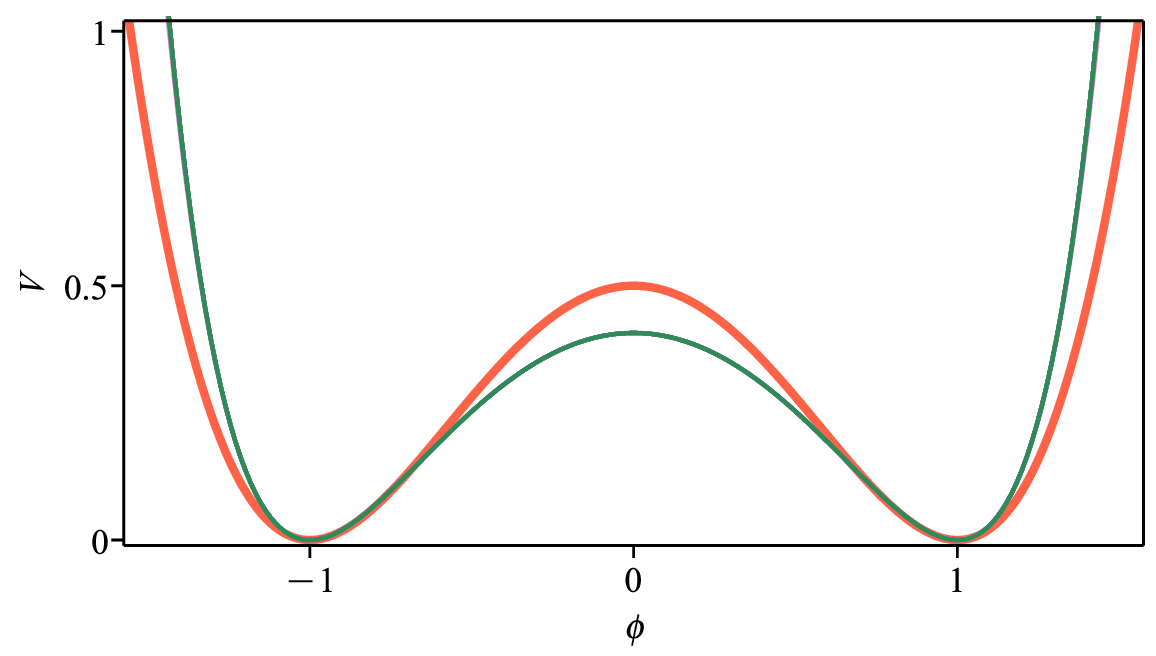}
    \includegraphics[width=8cm]{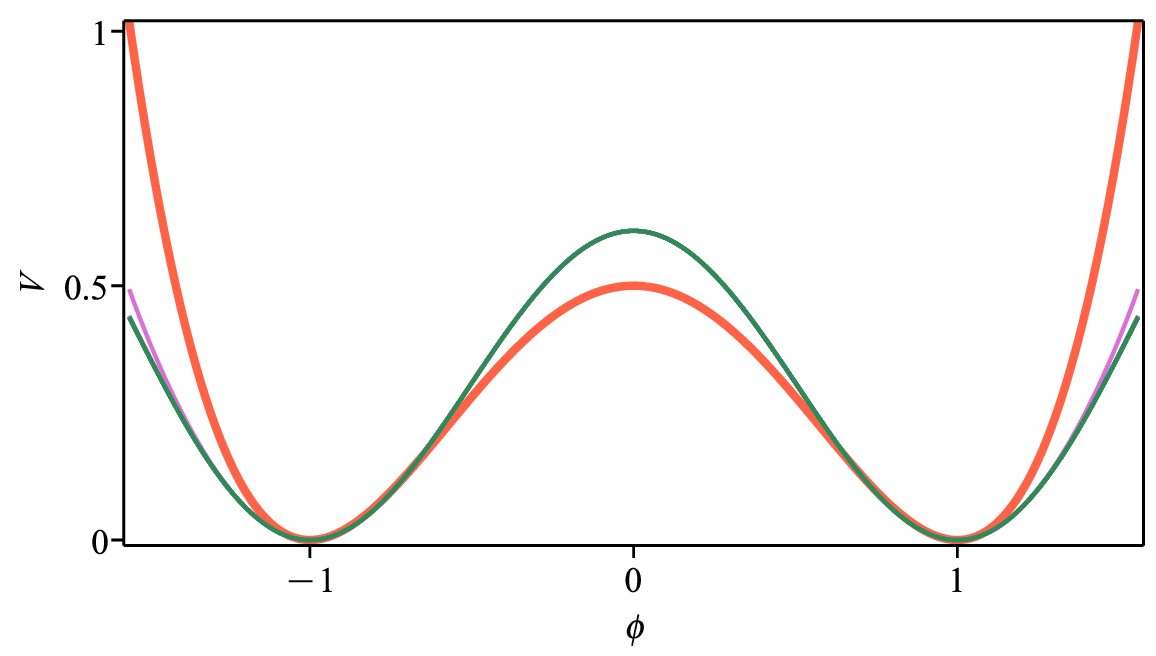}
    \includegraphics[width=8cm]{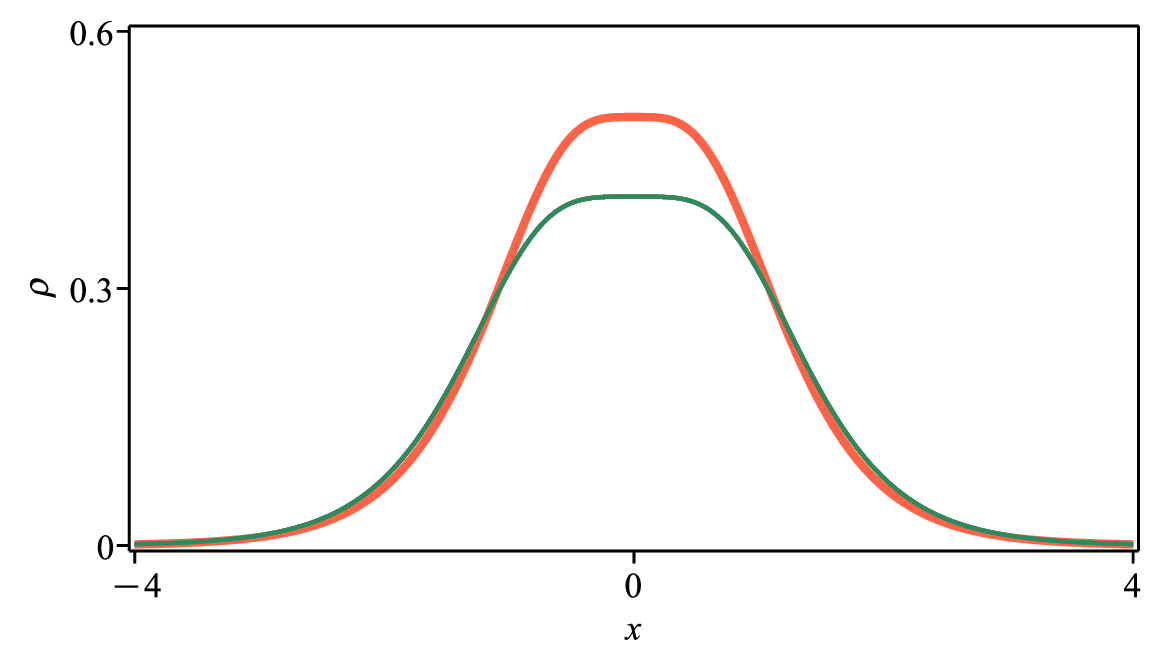}
    \includegraphics[width=8cm]{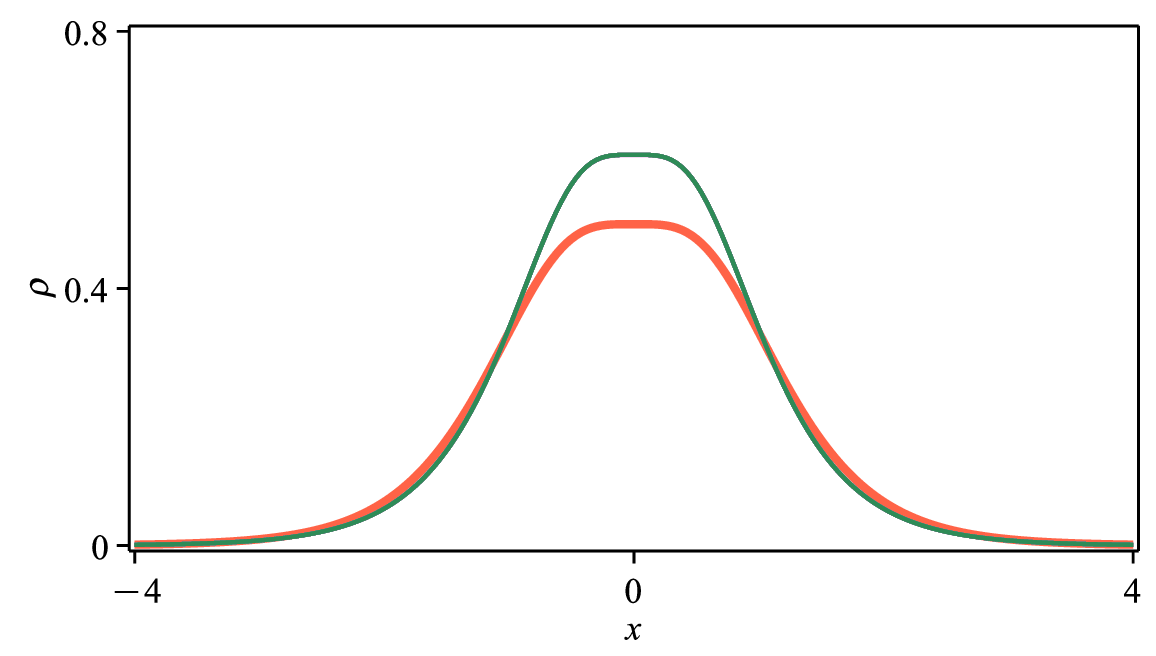}
    \caption{Same as Fig.~\ref{fig3} except that for the left side it is for the coefficient $c_j=1/(2j)!$, and for the right side, $c_j=(-1)^j/(2j)$.}
\label{fig5}    
\end{figure}

\begin{figure}[htb!]
    \centering
    \includegraphics[width=8cm]{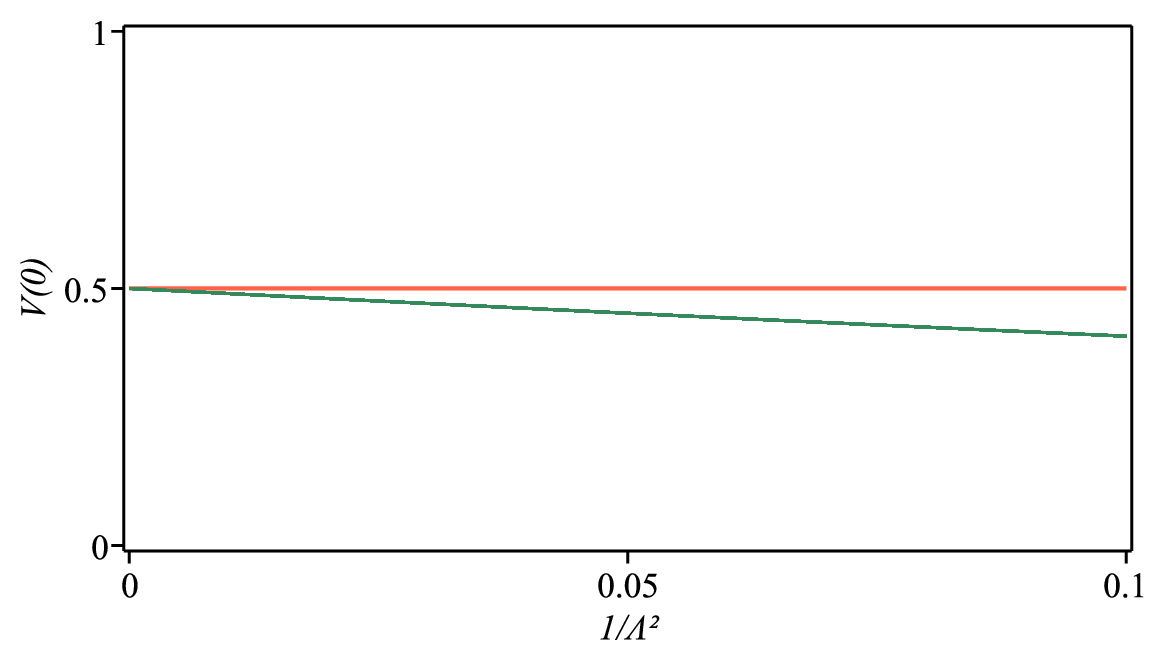}
    \includegraphics[width=8cm]{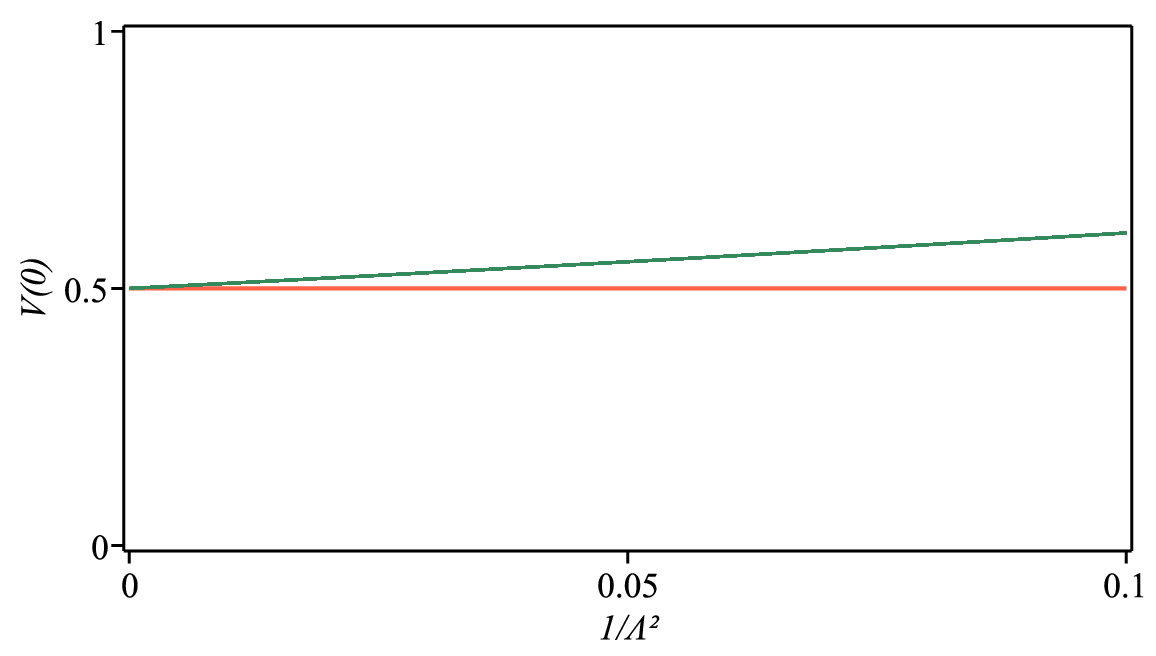}
    \includegraphics[width=8cm]{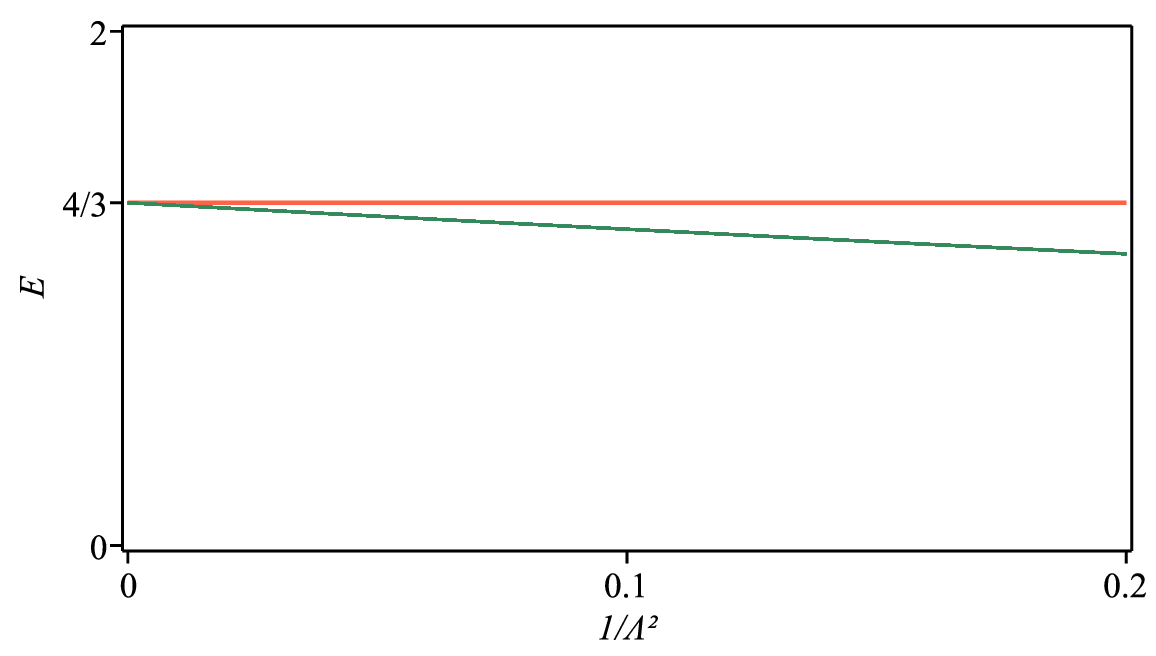}
    \includegraphics[width=8cm]{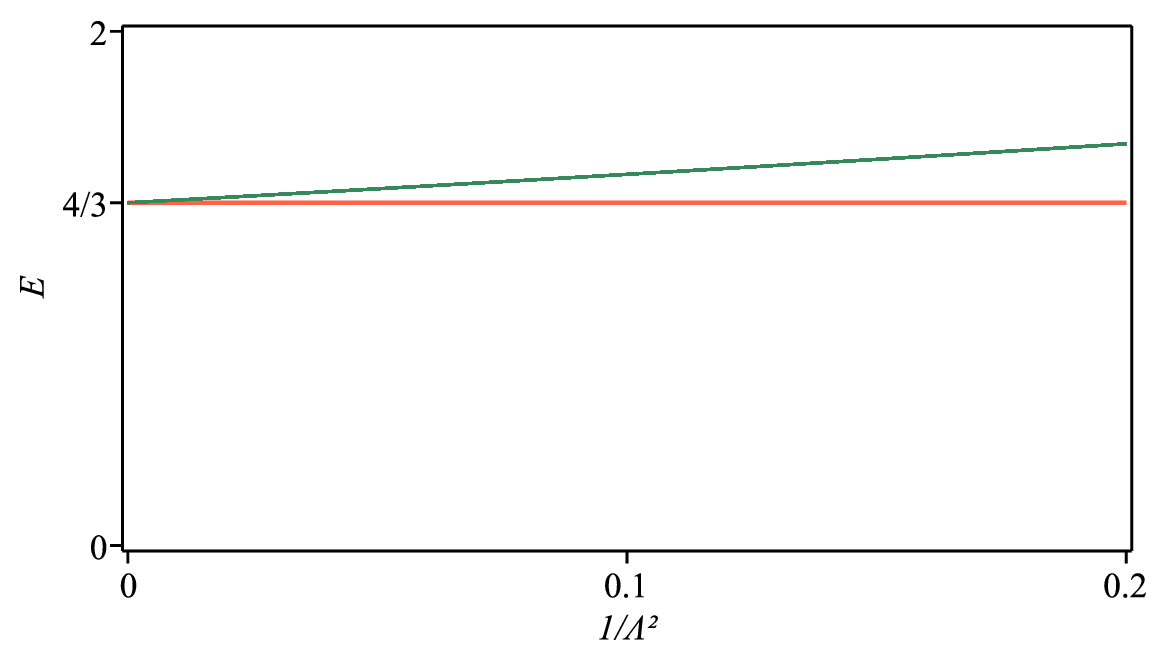}
    \caption{Same as Fig.~\ref{fig4} except that for the left side it's for the coefficient $c_j=1/(2j)!$, and for the right side, $c_j=(-1)^j/(2j)$.}
\label{fig7}    
\end{figure}

Now, we examine two examples where $\lim\limits_{j\to\infty}K_j/\Lambda^{2j}=0$ and $\lim\limits_{j\to\infty}\Ec_j/\Lambda^{2j}=0$ are convergent. With this goal, we consider the form factors in Eqs.~\eqref{f3} and \eqref{f4}. The coefficients of these functions are given by $c_j=1/(2j)!$ and $c_j=(-1)^j/(2j)!$, and the potential height \eqref{V0} is written as
\be
V(0) = \frac{\lambda v^4}{2} \mp\frac{\lambda^2 v^6}{\Lambda^2} +\frac{3\lambda^3 v^8}{4\Lambda^4} \mp\frac{19\lambda^4 v^{10}}{45\Lambda^6} +\frac{269\lambda^5 v^{12}}{1260\Lambda^8} \mp\frac{487\lambda^6 v^{14}}{4725\Lambda^{10}} +\mathcal{O}\left(\frac{1}{\Lambda^{12}}\right).
\ee
In this case, we require $\Lambda^2>2\lambda v^2$, so that its height is not too different from $V(0)=\lambda v^4/2$. The expansion of the potential from Eq.~\eqref{Vgeral} and the respective energy density \eqref{rho} can be seen in Fig.~\ref{fig5} for $\lambda=v=1$ and $\Lambda^2=10$. In the top panels, one observes that the shape of potential is similar to the local $\phi^{4}$-potential, regardless of the order of $1/\Lambda^2$. We also check its derivatives $V_\phi$, $V_{\phi\phi}$, and $V_{\phi\phi\phi}$, and  their behavior does not differ essentially. The energy densities in the bottom panels are displayed, which have a similar shape to Eq.~\eqref{rho0}. Both the potentials $V(\phi)$ and the energy densities $\rho(x)$ change in their height, which depends on the value of $\Lambda$. Their height approach to the local case when $1/\Lambda^2$ tends to be zero; see the top panels of Fig.~\ref{fig7}. Note that there are no divergences for large values of $\Lambda$, which is different from what happens in Fig.~\ref{fig4}. We also observe that in the bottom panels of Fig.~\ref{fig7}, the energy per unit area $E$ depends on the typical high energy scale similar to the potential height.

We see that the limits $\lim\limits_{j\to\infty}K_j/\Lambda^{2j}=0$ and $\lim\limits_{j\to\infty}\Ec_j/\Lambda^{2j}=0$ are ensured for the coefficients in \eqref{f3} and \eqref{f4}, as they decay faster than those in \eqref{f1} and \eqref{f2}. Therefore, not every form factor satisfies this requirement. However, we tested some others whose coefficients are: $c_j=1/(j^2)!$, $c_j=1/(j^j)!$, $1/((j)!^3)$, $1/((j)!^j)$, $1/(j^8(j)!)$, $1/(j^j(j)!)$, and so on.

\section{Summary and conclusions}
In this work we have investigated domain wall solution in the context of a nonlocal real scalar field, whose action is given by Eq.~\eqref{model}. In this situation, we have seen that its equation of motion and energy density are very different from the local case for static configurations, containing infinite-order differential operators. We consider a general form factor to extend the result from the Ref.~\cite{nonlocaldefects}. To illustrate our results, in addition to the previously studied case \eqref{f2}, we chose other form factors: the exponential \eqref{f1}, hyperbolic \eqref{f3}, and oscillatory \eqref{f4} functions. 

We split our investigation into three different cases: first, we consider the simplest potential $\phi^2$, leading to a nonlocal version of the standard Klein-Gordon equation \eqref{xeomKG}. The solutions for this case are useful for understanding the behavior far from the origin of domain wall solutions. Compared to the local solution, we observe that the exponential behavior remains, going faster or slower toward the vacuum, depending on the choice of the form factor. The second case consists of choosing the $\phi^4$ potential and applying the perturbation scheme used by the authors \cite{nonlocaldefects} in order to get topological solutions. We found general results regardless of the choice of the form factor for expansion up to $1/\Lambda^2$, including the particular form factor used in the aforementioned reference. Additionally, we note that when it is imposed that the perturbed solution maintains the monotonicity of the local solution, the expansion coefficient $c_1$ must be positive. We also specifically investigated three regions of the solution: i) near the origin $|x|<x_1$; ii) far from the origin $|x|>x_2$; iii) the intermediate region $x_1\leq |x|\leq x_2$. We verified for a generic form factor that the perturbative method is not valid for the region (iii), as it was previously found for a specific form factor \cite{nonlocaldefects}. To shed more light on this, we explicitly calculated the second-order contribution of the perturbation scheme (up to $1/\Lambda^4$). Note that the intermediate region, where the method breaks down, has expanded its range ($\tilde{x}_1\leq |x|\leq \tilde{x}_2$ with $\tilde{x}_1<x_1$ and $\tilde{x}_2>x_2$). However, it is important to emphasize that the results near and far from the origin still hold, reinforcing the previous result. 

Finally, we addressed the problem from another perspective. We have assumed an analytical expression for the scalar field solution: $\phi(x)=v\tanh\big(\sqrt{\lambda}vx\big)$, and then we substituted it into the equation of motion \eqref{xeom} to obtain the potential $V(\phi)$ that depends on the typical high energy scale. The potential has been expressed as a sum of infinite terms and is expected to differ slightly from the potential of the local model ($\phi^4$). For this, we notice that its convergence depends directly on the coefficients $c_j$ of the form factor. A similar analysis was also conducted for energy per area of the solution, and we obtained the same results regarding convergence.

For future investigations, we could explore nonlocal scalar field models in curved space-time in different contexts, such as the braneworld scenario whose various aspects were considered in \cite{wolfe,gremm1,gremm2,trilogia3,bentBI} and hyperscaling violating geometries defined in \cite{lifs2,lifs3}. Another possible direction involves studying collisions, as explored in \cite{colisao1,colisao2,colisao3,colisao4,colisao5}. We also plan to extend this study by coupling it with gauge field,  consider the dependence on two spatial coordinates, and search for vortex solutions  earlier discussed in the context of local field theories, f.e. in \cite{vortex1,vortex2,vortex3,vortex4,vortex5}. We also intend to apply this methodology to investigating exact solutions, in particular, those ones describing compact objects in nonlocal gravity theories. 

\appendix

\section{Perturbation around the local solution}\label{pertub}
Inspired by Ref.~\cite{nonlocaldefects}, we develop a perturbative method for nonlocal solutions. We consider a perturbation $\xi_\Lambda$ around the local solution $\phi_0$ of the type
\be\label{solpertubGERAL}
\phi(\Vec{r},t) = \phi_0(\Vec{r},t) +\xi_\Lambda(\Vec{r},t).
\ee
The perturbation depends on the typical high energy scale, and we require that this contribution vanishes in the limit $\Lambda \to \infty$. This perturbation is expected to be small compared to the local solution, that is, $|\phi_0| \gg |\xi_\Lambda|$. Using the equation of motion \eqref{feom}, we have that
\be\label{eompert}
\Box f(\Box_\Lambda)\phi_0 +\Box f(\Box_\Lambda)\xi_\Lambda = V_\phi.
\ee
Considering that the local solution satisfies $\Box\phi_0=V_\phi^{(0)}$, we rewrite the above equation as
\be
f(\Box_\Lambda)V_\phi^{(0)} +\Box f(\Box_\Lambda)\xi_\Lambda = V_\phi.
\ee
Expanding the nonlocal terms using Eq.~\eqref{fexp}, we obtain
\bes
\bal
f(\Box_\Lambda)V_\phi^{(0)} &= V_\phi^{(0)} +\frac{c_1}{\Lambda^2}\frac{d^2V_\phi^{(0)}}{dx^2} +\mathcal{O}\left(\frac{1}{\Lambda^{4}}\right),\\
f(\Box_\Lambda)\xi_\Lambda &= \xi_\Lambda +\frac{c_1}{\Lambda^2}\frac{d^2\xi_\Lambda}{dx^2} +\mathcal{O}\left(\frac{1}{\Lambda^{4}}\right).
\eal
\ees
We want these expansions to satisfy the following conditions:
\be\label{condsGERAL}
\left|V_\phi^{(0)}\right| \gg \left|\frac{c_1}{\Lambda^2}\frac{d^2V_\phi^{(0)}}{dx^2}\right| \quad\text{and}\quad |\xi_\Lambda| \gg \left|\frac{c_1}{\Lambda^2}\frac{d^2\xi_\Lambda}{dx^2}\right|,
\ee
as well as the consecutive terms, as long as they follow the hierarchy of orders in $1/\Lambda^2$.

Our next step is to develop this only for static and one-dimensional solutions, $\phi=\phi(x)$, to the order of $1/\Lambda^2$. Writing the perturbation $\xi_\Lambda$ as
\be
\xi_\Lambda = \sum_{j=1}^n\frac{\xi_j}{\Lambda^{2j}},
\ee
It is possible to expand up to the desired order $n$. The lowest order is when $n=1$, which is the initial case considered. Thus, the first-order approximation is
\be\label{phi1GERAL}
\phi_1(x) = \phi_0(x)+\frac{1}{\Lambda^2}\xi_1(x).
\ee
Using Eq.~\eqref{eompert}, the contribution $1/\Lambda^2$ leads to a non-homogeneous second-order ODE, given by
\be\label{edo1o}
\frac{d^2\xi_1}{dx^2} +c_1\frac{d^4\phi_0}{dx^4} = V_{\phi\phi}|_{\phi_0}\xi_1,
\ee
with solution
\be
\xi_1(x) = \left(\Nc_1+\Nc_2\Ic(x) +c_1\left(\frac12\bigg(\frac{d^2\phi_0}{dx^2}\bigg)^{\!2}-\frac{d\phi_0}{dx}\frac{d^3\phi_0}{dx^3}\right)\Ic(x) +c_1\!\int^x\! dy\left(\Ic(y)\frac{d\phi_0}{dy}\frac{d^4\phi_0}{dy^4}\right)\!\right)\!\!\bigg(\frac{d^2\phi_0}{dx^2}\bigg),
\ee
where $\Ic(x) = \int^x dy\,\left(d\phi_0/dy\right)^{-2}$. Although we have the general solution for the perturbation $\xi_1$, solving the integrals analytically is not always feasible.

Simply obtaining the solution, however, is not sufficient; the validity of the conditions \eqref{condsGERAL} and the one below Eq.~\eqref{solpertubGERAL} must also be verified. To do so, we define two functions, $h(x)$ and $H(x)$, as follows
\be\label{hH}
h(x) = \frac{1}{\Lambda^2}\left|\frac{\xi_1(x)}{\phi_0(x)}\right| \ll 1 \quad\text{and}\quad H(x) = \frac{1}{|\xi_1|}\left|\frac{c_1}{\Lambda^2}\frac{d^2\xi_1}{dx^2}\right| \ll 1.
\ee
In general, these conditions cannot be satisfied over the entire range of the solution. The first condition, analyzed through the function $h(x)$, is typically easier to evaluate, as it can be adjusted by choosing appropriate values for $\Nc_1$ and $\Nc_2$. In contrast, the second condition, involving $H(x)$, is more challenging. To approach this, we use Eq.~\eqref{edo1o} to rewrite the second condition as
\be
\frac{1}{\xi_1}\frac{d^2\xi_1}{dx^2} = \frac{d^3\phi_0}{dx^3}\left(\frac{d\phi}{dx}\right)^{-1} -\frac{c_1}{\xi_1}\frac{d^4\phi_0}{dx^4}.
\ee
In most cases, the first term on the right-hand side of the above equation does not present issues, whereas the second term is more cumbersome. One way to avoid possible divergences in the function $H(x)$ is to find a local scalar field model where its fourth derivative, $d^4\phi_0/dx^4$, compensates for any zeros in the perturbation $\xi_1$. Beyond the local solution $\phi_0$ in Eq.~\eqref{kink}, it is possible to obtain the perturbation $\xi_1$ analytically for other kink-type solutions, such as Sine-Gordon, $\lambda\phi^6$, Christ-Lee, and long-range models \cite{sinegordon,lohe,tdlee,longrange}. Unfortunately, there are always certain intervals in parameter space where the perturbation theory breaks down.

The energy density of the solution \eqref{phi1GERAL} is given by
\be
\rho(x) = \rho_0(x) +\frac{1}{\Lambda^2}\rho_1(x),
\ee
where
\bes
\bal
\rho_0(x) &= -\frac12\phi_0\frac{d^2\phi_0}{dx^2} +V(\phi_0),\\
\label{rho1}\rho_1(x) &= -\frac{c_1}{2}\phi_0\frac{d^4\phi_0}{dx^4} -\frac{1}{2}\xi_1\frac{d^2\phi_0}{dx^2} -\frac{1}{2}\phi_0\frac{d^2\xi_1}{dx^2} +V_{\phi}|_{\phi_0}\xi_1,
\eal
\ees
are, respectively, the local energy density ($\rho_0$) and the first-order $1/\Lambda^2$ contribution ($\rho_1$).

We can also consider higher orders of $1/\Lambda^2$. For second-order corrections ($n=2$), the solution is  $\phi_2(x)=\phi_1(x)+(1/\Lambda^4)\xi_2(x)$, which is determined by solving
\be\label{edo2o}
\begin{aligned}
\frac{d^2\xi_2}{dx^2} +c_1\frac{d^4\xi_1}{dx^4} +c_2\frac{d^6\phi_0}{dx^6} = \frac{1}{2}V_{\phi\phi\phi}|_{\phi_0}\xi_1^2 +V_{\phi\phi}|_{\phi_0}\xi_2.
\end{aligned}
\ee
The validity of the perturbation must also be verified using the conditions below Eq.~\eqref{solpertubGERAL} and in Eq.~\eqref{condsGERAL}, considering terms up to $1/\Lambda^4$. The correction of the energy density $\rho_2(x)$ can be obtained from the above solution. This analysis can be extended to the $n$-th order, provided the preceding $n-1$ orders have been obtained.

\acknowledgments{We acknowledge financial support from the Brazilian agencies Conselho Nacional de Desenvolvimento Cient\'ifico e Tecnol\'ogico (CNPq), Grants 310994/2021-7 and 402830/2023 (RM), 303777/2023-0
(AYP) and 307628/2022-1 (PJP), and Paraiba State Research Foundation (FAPESQ-PB), Grants 2783/2023 (IA), 0003/2019 (RM), 0015/2019 (AYP) and 150891/2023-7 (AYP and PJP).}



\begin{thebibliography}{99}
\bibitem{rodape} Indeed, the Einstein-Hilbert term can be dynamically induced in the low-energy limit of these theories when a field beyond the Standard Model—such as a scalar field \cite{Salvio} or a vector field \cite{Lehum}—acquires a nontrivial vacuum expectation value.
\end{thebibliography}

\end{document}